\documentclass[aps,prb,showpacs,a4paper,showkeys,amssymb,amsmath,
superscriptaddress, floatfix,
twocolumn]
{revtex4-1}
\usepackage{graphicx}
\usepackage{tabularx}
\usepackage{amssymb}
\usepackage{color}
\usepackage{hyperref}
\usepackage{bm}
\usepackage{dcolumn}

\renewcommand{\vec}[1]{\bm{#1}}
\newcommand{\mat}[1]{\underline{\bm{#1}}}

\begin{document}

\title{Brillouin zone unfolding method for effective phonon spectra}

\author{Timothy B. Boykin}
\email[]{boykin@ece.uah.edu}
\affiliation{Department of Electrical and Computer Engineering, 
The University of Alabama in Huntsville, Huntsville, Alabama 35899, USA}

\author{Arvind Ajoy}
\email[]{aajoy@nd.edu}
\affiliation{Network for Computational Nanotechnology, 
School of Electrical and Computer Engineering, Purdue University, West Lafayette,
Indiana 47907, USA}
\affiliation{Department of Electrical Engineering, 
University of Notre Dame, Indiana 46656, USA}

\author{Hesameddin Ilatikhameneh}
\email[]{hilatikh@purdue.edu}
\affiliation{Network for Computational Nanotechnology, 
School of Electrical and Computer Engineering, Purdue University, West Lafayette,
Indiana 47907, USA}

\author{Michael Povolotskyi}
\email[]{mpovolot@purdue.edu}
\affiliation{Network for Computational Nanotechnology, 
School of Electrical and Computer Engineering, Purdue University, West Lafayette,
Indiana 47907, USA}

\author{Gerhard Klimeck}
\email[]{gekco@purdue.edu}
\affiliation{Network for Computational Nanotechnology, 
School of Electrical and Computer Engineering, Purdue University, West Lafayette,
Indiana 47907, USA}

\date{\today}

\begin{abstract}
Thermal properties are of  great interest in modern electronic devices
and nanostructures.   Calculating these properties  is straightforward
when the device is made from  a pure material, but problems arise when
alloys are used. Specifically,  only approximate bandstructures can be
computed  for  random  alloys  and  most  often  the  Virtual  Crystal
Approximation  (VCA)  is  used.   Unfolding  methods  [T.  B.  Boykin,
N.   Kharche,   G.   Klimeck,    and   M.   Korkusinski,   J.   Phys.:
Condens.  Matt.  19,  036203  (2007).]  have proven  very  useful  for
tight-binding calculations  of alloy electronic  structure without the
problems   in  the   VCA,   and  the   mathematical  analogy   between
tight-binding and valence-force-field approaches to the phonon problem
suggest  they  be employed  here  as  well.  However, there  are  some
differences in the physics of the two problems requiring modifications
to  the electronic structure  approach. We  therefore derive  a phonon
alloy   bandstructure  (vibrational  mode)   approach  based   on  our
tight-binding    electronic    structure    method,   modifying    the
band-determination  method  to   accommodate  the  different  physical
situation. Using  the method, we study In$_x$Ga$_{1-x}$As  alloys and find
very good agreement with available experiments.
\end{abstract}

\maketitle

\section{Introduction}
\label{sec_Introduction}
Accurate  modeling of the  thermal properties  of semiconductors  is a
technologically   significant  problem:  Heat   degrades  conventional
transistor  performance  and   nanowires  are  becoming  important  in
next-generation          electronics         \cite{Lieber_NAture_2001,
Persson_NanoLett_2006}.  In  addition, altering phonon  properties via
isotopic alloy disorder has been investigated as a possible method for
improving    the    performance    of    carbon    nanotube    devices
\cite{Vandecasteele_PRL_2009}   and  high   power  GaN   field  effect
transistors  \cite{Khurgin_APL_2008}.    Modeling  thermal  properties
requires  in turn  accurate  phonon  spectra (or  bands),  which is  a
straightforward task  for pure materials  (Si, Ge, GaAs,  InAs, etc.).
Alloys,  both bulk and  nanostructure, are  used in  numerous advanced
devices and modeling their properties, both thermal and electronic, is
more  difficult  because  translational  symmetry exists  in  only  an
approximate sense. The simplest alloy treatment is the Virtual Crystal
Approximation (VCA), but it does not accurately capture the effects of
random alloying in electronic structure calculations for both bulk and
nanostructures  \cite{Boykin_TNano_2007, Kharche_JCE_2008}.   To date,
most thermal properties  modeling of alloys has been  with the VCA, as
in   the   In$_x$Ga$_{1-x}$As    bulk   and   nanowire   calculations   in
Ref.  \onlinecite{Salmani_JCE_2012}.  While VCA  phonon models  do use
realistic  underlying   models  such  as   valence  force-field  (VFF)
approaches, the VCA is expected to have similar deficiencies for these
cases   as  for  electronic   structure  calculations,   perhaps  even
worse. The reason for expecting  worse VCA phonon bands comes directly
from the  periodic table: Exchanging an  atom for another  in the same
column involves  a very large change  in mass, such as  Ga (69.72) vs.
In (114.8).  Conversely, in electronic structure calculations, such an
atomic exchange  generally results in  more modest overall  changes to
the inter- and intra-atomic parameters.

Electronic structure  calculations beyond the  VCA are often  based on
applying  Brillouin zone unfolding  to random-alloy  supercells, using
either    tight-binding    \cite{Boykin_JPCM_2007,    Boykin_PRB_2005,
Boykin_PRB_2007}     or     pseudopotential     \cite{Zunger_PRL_2010,
Zunger_PRB_2012} bases. Unfolding has also been applied to the complex
bands of surfaces \cite{Ajoy_JPCM_2012}  and several other variants of
the  method   have  been  proposed   \cite{Allen_PRB_2013,  *[Erratum:
]Allen_PRB_2013erratum,  Ku_PRL_2010, Berlijn_PRB_2014, Lee_JPCM_2013,
Huang_NJP_2014,  Medeiros_PRB_2014,   Peil_PRB_2012}.   These  methods
create a  random alloy  supercell (SC) having  a very large  number of
primitive  cells  (PCs),  then  unfold  the  supercell  bands  onto  a
primitive-cell  periodic  basis.   Sum  rules  \cite{Boykin_JPCM_2007}
allow one  to define  average energies and  approximate bands.   For a
sufficiently large supercell, the effects of random alloying should be
captured in the effective bandstructure.  Supercell-unfolded effective
bandstructures  have  many  advantages  over  VCA  bands  because  the
unfolded bands reproduce trends which  the VCA cannot, such as bandgap
bowing in AlGaAs \cite{Boykin_JPCM_2007}.

This  success of  supercell-unfolding effective  bandstructure methods
for electronic structure calculations argues strongly that they should
be applied  to the problem of  alloy phonon spectra as  well. To date,
most  applications of  unfolding to  vibrational problems  has  been to
simple one- dimensional problems  which do not encompass random alloys
\cite{Allen_PRB_2013}.  Other supercell effective phonon bandstructure
methods beyond  the VCA involve  creating a phase-  and force-constant
averaged  primitive-cell dynamical  matrix  \cite{Wang_JPCM_2011}.  In
this    work    we     modify    our    supercell-unfolding    methods
\cite{Boykin_JPCM_2007, Boykin_PRB_2005,  Boykin_PRB_2007} to optimize
them for the alloy phonon problem, studying the behavior of the phonon
bands of In$_x$Ga$_{1-x}$As as  a function of mole fraction, comparing
the results to the VCA, and calculating the sound velocity.  The paper
is organized  as follows: Sec.  \ref{sec_Method}  presents the method;
Sec. \ref{sec_Results} the results; and Sec. \ref{sec_Conclusions} the
conclusions.

\section{Method}
\label{sec_Method}
\subsection{Allowed primitive-cell wavevectors}
\label{sec_AllowedPCs}
The  SC bands  will be  unfolded onto  a PC-periodic  basis.  A  PC is
defined  by direct translation  vectors, $\vec{a}_j,  j =  1,2,3$, not
necessarily orthogonal;  the corresponding reciprocal  lattice vectors
of the  primitive cell  are denoted $\vec{b}_j,  j =  1,2,3$. Born-von
Karman boundary conditions are imposed  over the SC, which is composed
of $N_j$ primitive cells in  the $\vec{a}_j$ direction, for a total of
$N_c = N_1  N_2 N_3$ PCs.  SC states of  SC wavevector $\vec{Q}$ (this
implies the existence of $N_s$  SCs over which further Born-von Karman
boundary  conditions  are  imposed)   unfold  onto  PC  states  of  PC
wavevector, $\vec{q}_m$:
\begin{align}
\label{eq_q}
\vec{q}_m = \vec{Q} + \vec{G}_m, m = 1, 2, \dots, 3.
\end{align}
The SC reciprocal lattice vectors $\vec{G}_m$ are :
\begin{align}
\label{eq_G}
\vec{G}_m & = \sum_{j=1}^3 \frac{n_j}{N_j} \vec{b}_j, \\
\nonumber n_j & = 
\begin{cases}
-(N_j - 2)/2, \ldots, 0, \ldots, N_j/2, & N_j \text{ even} \\
-(N_j - 1)/2, \ldots, 0, \ldots, (N_j - 1)/2, & N_j \text{ odd}
\end{cases}
\end{align}
In eq. (\ref{eq_G}) the index m  corresponds to one of the $N_c$ trios
$(n_1, n_2,  n_3)$, and  if any $\vec{q}_m$  falls outside the  PC first
Brillouin Zone it  is translated back in by  adding the appropriate PC
reciprocal     lattice      vector.      Our     previous     software
\cite{Boykin_JPCM_2007,   Boykin_PRB_2005,  Boykin_PRB_2007}  required
rectangular SCs,  which necessitated using non-  primitive small cells
and       hence      additional      allowed       PC      wavevectors
\cite{Boykin_PhysicaE_2009}.     Our    new    version    accommodates
non-rectangular  SCs, thus  allowing us  to avoid  these complications
\cite{Steiger_TNano_2011, Fonseca_JCE_2013}.

\subsection{Primitive Cells}
\label{sec_PrimitiveCells}
For probing the bands in  the $[100]$, $[110]$, and $[111]$ directions
we  use  different PCs  for  zincblende.   The  direct and  reciprocal
lattice     vectors    for    the     $[lmn]$    PC     are    denoted
$\vec{\alpha}_j^{[lmn]}$    and    $\vec{\beta}_j^{[lmn]}$, 
respectively.  The   specific  definitions  are   given  in  Cartesian
coordinates in Table  I. Note that all three cells  defined in Table I
are  indeed  primitive, since  for  all three  $\vec{\alpha}_1^{[lmn]}
\cdot  (  \vec{\alpha}_2^{[lmn]}  \times  \vec{\alpha}_3^{[lmn]}  )  =
a^3/4$. An $[lmn]$-SC  has a large value for $N_1$ so  as to probe the
PC   bands  with   a  very   fine  resolution   along   the  direction
$\vec{\beta}_1^{[lmn]}$.  Aravind \cite{Aravind_AJP_2006}   gives  a  general
method for  determining the $\vec{\alpha}_j$ .  The Appendix discusses
the  portions of  the PC  Brillouin zone  probed by  calculations using
these cells.

\begin{table*}[t]
  \caption{Cartesian coordinates $(x, y, z)$ of direct and reciprocal PC lattice vectors used.
  Units of $\vec{\alpha}_j^{[lmn]}$ are $a/2$, and units of $\vec{\beta}_j^{[lmn]}$ are $2 \pi /a$,
  where a is the conventional FCC cube edge.}
  \begin{ruledtabular}
    \begin{tabular}{ccccccc}
      &  
      $\vec{\alpha}_1^{[lmn]}$  & 
      $\vec{\alpha}_2^{[lmn]}$  &
      $\vec{\alpha}_3^{[lmn]}$  &
      $\vec{\beta}_1^{[lmn]}$  &
      $\vec{\beta}_2^{[lmn]}$  &
      $\vec{\beta}_3^{[lmn]}$   \\
      \colrule 

      $[100]$  & 
      $(1,1,0)$& 
      $(0,1,1)$&
      $(0,-1,1)$&
      $(2,0,0)$ &
      $(-1,1,1)$ &
      $(1,-1,1)$\\

      $[110]$ &
      $(1,0,1)$&
      $(1,-1,2)$&
      $(1,-1,0)$&
      $(2,2,0)$&
      $(-1,-1,1)$&
      $(1,-1,-1)$\\

      $[111]$ &
      $(1,1,0)$&
      $(0,-1,1)$&
      $(1,0,-1)$&
      $(1,1,1)$&
      $(1,-1,1)$&
      $(1,-1,-1)$\\
    \end{tabular}
  \end{ruledtabular}
  \label{tab_Cells}
\end{table*}

\subsection{Unfolding applied to Supercells}
The   phonon  unfolding  problem   is  mathematically   equivalent  to
electronic structure unfolding  when an underlying tight-binding basis
is used. In  the phonon case, each atom has  three degrees of freedom,
$x$, $y$, and $z$, which play  the role of orbitals in a tight-binding
model. That  is, when the problem  is written in  matrix notation, the
components of the motion for an  atom appear exactly as do orbitals in
a   tight-binding  model.   As  in   electronic   structure  unfolding
\cite{Boykin_JPCM_2007,  Boykin_PRB_2005, Boykin_PRB_2007}  we express
the  states in  terms of  SC-  and PC-periodic  basis functions,  then
observe  that a  SC state  of wavevector  $\vec{Q}$ must  be  a linear
combination  of the  PC states  of  wavevectors $\vec{q}_m  = \vec{Q}  +
\vec{G}_m$,   $m  =   1,2,  \dots,   N_c$.   Unfolding   recovers  the
contribution of  all PC states of  wavevector $\vec{q}_m$ to  a given SC
state   of  wavevector   $\vec{Q}$,  and   sum  rules   allow  PC-band
determination.

The phonon  spectra calculation is treated in  standard references and
texts;   our  notation   and  treatment   follows  that   of  Madelung
\cite{Madelung_Book_1981}.  First, we  consider the  case  of Born-von
Karman    boundary   conditions   applied    a   single    SC   (i.e.,
$\vec{Q}=\vec{0}$), consisting of  $N_c = N_1 N_2 N_3$  PCs.  Here the
normal mode amplitudes are written:
\begin{align}
\label{eq_normalmode}
u_{n, \alpha, l} = b^{(\alpha, l)} \exp( \iota \vec{q} \cdot \vec{\rho}_n)
\end{align}
where $n$ is the  primitive-cell index; $\vec{\rho}_n$ is the location
of the $n$-th PC relative to the SC origin; $\alpha$ is the atom in the
primitive-cell ($\alpha  = 1, 2, \ldots  , r$ where each  cell has $r$
atoms),  and   $l  =  x,  y,   z$  is  the   Cartesian  coordinate  of
motion. $\vec{q}$ is the  PC phonon wavevector. Because PC periodicity
is  enforced, modes  of different  $\vec{q}$ decouple  and  the $s$-th
eigenstate (of $3r$ total) satisfies the Hamiltonian matrix equation:
\begin{align}
\label{eq_eigenproblem}
\omega_s^2(\vec{q}) b_s^{(\alpha, l)} & = \sum_{\alpha', l'}
                         D_{(\alpha, l), (\alpha', l')}(\vec{q})
                         b_s^{(\alpha', l')}, \\
\nonumber
s & = 1,2, \dots, 3r,                                     
\end{align}
where  the  dynamical matrix  $\mat{D}$,  is  Hermitian and depends on 
the ion-ion interaction, $V_{ion-ion}$:
\begin{align}
\label{eq_dynamicalmatrix}
\nonumber
D_{(\alpha, l), (\alpha', l')}(\vec{q}) & =  \frac{1}{\sqrt{M_{\alpha} M_{\alpha'}}} 
                       \times \\
                     & \quad \sum_m \Phi_{(\alpha, l), (\alpha', l')}(m)
                             \exp(-\iota \vec{q} \cdot \vec{\rho}_m) \\
\label{eq_forceconstant}
\Phi_{(\alpha, l), (\alpha', l')}(m)& = \dfrac{ \partial^2 V_{ion-ion}}
                     { \partial \rho_{0, \alpha, l} \partial \rho_{m, \alpha', l'} }
\end{align}
The eigenvector  for the $s$-th  eigenstate $s =  1, 2, \dots  ,3r$ is
written as a column vector
\begin{align}
\label{eq_b}
\vec{b}_s = \left. \begin{bmatrix}
              b_s^{(1,x)} \\
              b_s^{(1,y)} \\
              \vdots \\
              b_s^{(r,z)}
            \end{bmatrix} \right\} 
            \text{ 3$r$ rows, $r$ = atoms / PC } 
\end{align}
Because the eigenproblem,  eq. (\ref{eq_eigenproblem}) is Hermitian, the
$\vec{b}_s$ are orthonormal
\begin{align}
\label{eq_borthonormal}
\vec{b}_{s'}^{\dagger} \cdot \vec{b}_s = \delta_{s',s}, \quad s',s = 1,2,\ldots, 3r.
\end{align}

Next we consider the case of  $N_s$ SCs, for a total of $N_sN_1N_2N_3$
PCs, but continue to  enforce Born-von Karman boundary conditions over
a single SC.  Here each  amplitude just acquires an extra phase factor
based on the SC location,  $\vec{R}_j$ .  The amplitude for the $n$-th
PC in the $j$-th SC is
\begin{align}
\label{eq_u1}
\begin{split}
\vec{u}_{s,j}^{(n)} (\vec{Q} + \vec{G}_m) 
     = & \exp(\iota \vec{Q} \cdot \vec{R}_j ) \cdot \\
    &\exp{[\iota (\vec{Q} + \vec{G}_m) \vec{\rho}_n]}
    \cdot \vec{b}_s(\vec{Q} + \vec{G}_m) 
\end{split}
\end{align}
where  we  recall  from  subsection  \ref{sec_AllowedPCs}  above  that
$\vec{G}_m \cdot \vec{R}_j = 2  \pi \times \texttt{ integer}$. Now the
vector for the $j$-th supercell is
\begin{align}
\label{eq_u2}
\vec{u}_{s,j}(\vec{Q} + \vec{G}_m) = \frac{1}{\sqrt{N}_c}
                    \left.
                    \begin{bmatrix}
                       \vec{u}_{s,j}^{(1)} (\vec{Q} + \vec{G}_m) \\
                       \vdots \\
                       \vec{u}_{s,j}^{(N_c)} (\vec{Q} + \vec{G}_m) 
                    \end{bmatrix}
                    \right\}
                    \text{ $3rN_c$ rows.}
\end{align}
These    SC   vectors    remain   orthogonal:
\begin{align}
\label{eq_uorthonomal}
\begin{split}
\vec{u}_{s',j}^{\dagger}( \vec{Q} + \vec{G}_m ) 
& \cdot 
\vec{u}_{s,j}( \vec{Q} + \vec{G}_m)   = \\
\frac{1}{N_c}
&\sum_{n=1}^{N_c}\vec{u}_{s',j}^{(n) \dagger} (\vec{Q} + \vec{G}_m)  \cdot
              \vec{u}_{s,j}^{(n)} (\vec{Q} + \vec{G}_m) =  \\
\frac{1}{N_c}&\sum_{n=1}^{N_c} \delta_{s',s} = \delta_{s's}.
\end{split}
\end{align}
Finally,  when Born-von  Karman boundary  conditions are  only applied
over the entire set of SCs, the dynamical matrix eigenvalue problem is
now of  dimension $3rN_c$ and  the only wavevector  is that of  the SC
first  Brillouin   zone,  $\vec{Q}$;   for  each  there   are  $3rN_c$
eigenstates. The  $p$-th SC  eigenstate in the  $j$-th SC,  in analogy
with eq. (\ref{eq_q}) is written,
\begin{align}
\label{eq_V}
\nonumber
\vec{V}_{p,j}  & = \exp(\iota \vec{Q} \cdot \vec{R}_j) \times
            \left.
            \begin{bmatrix}
               \vec{v}_p^{(1)} \\
               \vdots \\
               \vec{v}_p^{(N_c)}
            \end{bmatrix} 
            \right\}
            \text{ $3rN_c$ rows, } \\
\vec{V}_{p',j}^{\dagger} \cdot \vec{V}_{p,j}  
& = \delta_{p,p'} \quad{p,p' = 1,2, \ldots, 3r N_c } \\
\label{eq_v}
\text{ where } \vec{v}_p^{(n)} & = \left.
                   \begin{bmatrix}
                     \beta_p^{(n,1,x)} \\
                     \vdots \\
                     \beta_p^{(n,r,z)} 
                   \end{bmatrix}
                   \right\}
                   \text{ $3r$ rows} 
\end{align}

The SC eigenvector of wavevector $\vec{Q}$ is generally a superposition of the
$N_c$ PC eigenvectors of wavevectors $\vec{q}_m = \vec{Q} + \vec{G}_m, m
= 1, 2,\ldots, N_c$. This  relationship is exact for perfect unfolding
and in  the case of imperfect  unfolding leads to a  useful ansatz for
determining approximate band edges.

The   unfolding  proceeds   as  in   the  electronic   structure  case
\cite{Boykin_JPCM_2007,  Boykin_PRB_2005, Boykin_PRB_2007}. Expressing
the SC vector as a superposition,
\begin{align}
\label{eq_projection1}
\vec{V}_{p,j} & = \sum_{s=1}^{3r} \sum_{m=1}^{N_c} 
             a_{p; (m,s)} \vec{u}_{s,j}(\vec{Q} + \vec{G}_m),  
\end{align}
then  using  eqs. (\ref{eq_u2})  and  (\ref{eq_V})  and selecting  the
$n$-th row of the system of equations yields:
\begin{align}
\label{eq_projection3}
\vec{v}_p^{(n)} = \dfrac{\exp(-\iota \vec{Q} \cdot \vec{R_j})}{ \sqrt{N_c}}
                \sum_{s=1}^{3r} \sum_{m=1}^{N_c} 
                a_{p; (m,s)} \vec{u}_{s,j}^{(n)}(\vec{Q} + \vec{G}_m).
\end{align}
Substituting      eq.     (\ref{eq_v})      on     the      LHS     of
eq.  (\ref{eq_projection3}),   eq.  (\ref{eq_u1})  on   the  RHS,  and
selecting the row corresponding to the $\alpha$-th basis atom, $\alpha
= 1, 2,\dots,  r$ , and the  $w$-th component of motion, $w  = \{x, y,
z\}$, one finds:
\begin{align}
\begin{split}
\label{eq_simplification2}
\exp(-\iota \vec{Q}\cdot\vec{\rho}_n) \beta_p^{(n, \alpha, w)} = &
 \sum_{m=1}^{N_c} \frac{1}{\sqrt{N_c}} \exp(\iota \vec{G}_m \cdot \vec{\rho}_n)
\times \\
& \quad \left[\sum_{s=1}^{3r}  a_{p; (m,s)}  b_s^{(\alpha, w)}(\vec{Q}+\vec{G}_m) \right]
\end{split}
\end{align}
Eq. (\ref{eq_simplification2})  is easily rearranged into  a system of
equations, coupling all $N_c$ of the PC states:
\begin{align}
\label{eq_mateqn1}
\vec{B}_p^{(\alpha, w)}(\vec{Q}) &= \mat{U} \cdot \vec{C}_p^{(\alpha, w)} (\vec{Q}), \\
\text{ with }
\nonumber\mat{U}^{\dagger} &= \mat{U}^{-1}
\end{align}
\begin{align}
\vec{C}_p^{(\alpha, w)}(\vec{Q}) &=
   \begin{bmatrix}
      \sum\limits_{s=1}^{3r}  a_{p; (1,s)}  b_s^{(\alpha, w)}(\vec{Q}+\vec{G}_1) \\
      \vdots \\
      \sum\limits_{s=1}^{3r}  a_{p; (N_c,s)}  b_s^{(\alpha, w)}(\vec{Q}+\vec{G}_{N_c}) \\
   \end{bmatrix}, \\
\mat{U}& = \frac{1}{\sqrt{N_c}}
   \begin{bmatrix}
    e^{\iota \vec{G}_1 \cdot \vec{\rho}_1} & \cdots  &  e^{\iota \vec{G}_{N_c} \cdot \vec{\rho}_1} \\
    e^{\iota \vec{G}_1 \cdot \vec{\rho}_2} & \cdots  &  \vdots \\
    \vdots & \ddots   &\vdots \\
    e^{\iota \vec{G}_1 \cdot \vec{\rho}_{N_c}} & \cdots &  e^{\iota \vec{G}_{N_c} \cdot \vec{\rho}_{N_c}} \\
  \end{bmatrix}, \\
\vec{B}_p^{(\alpha, w)}(\vec{Q}) &=
   \begin{bmatrix}
      e^{-\iota \vec{Q}\cdot\vec{\rho}_1} \beta_p^{(1, \alpha, w)} \\
      \vdots \\
      e^{-\iota \vec{Q}\cdot\vec{\rho}_{N_c}} \beta_p^{(N_c, \alpha, w)}
   \end{bmatrix}.
\end{align}
Because    $\mat{U}$   is    unitary,   we    can    trivially   solve
eq. (\ref{eq_mateqn1}) for $\vec{C}_p^{(\alpha, w)}(\vec{Q})$
\begin{align}
\label{eq_mateqn2}
\vec{C}_p^{(\alpha, w)}(\vec{Q}) = 
\mat{U}^{\dagger} \cdot \vec{B}_p^{(\alpha, w)}(\vec{Q}).
\end{align}
The $\vec{C}_p^{(\alpha,  w)}(\vec{Q})$ are the  quantities needed for
band determination, exact or approximate. 

\subsection{Sum Rule and Band Determination}
\label{sec_Sumrule}
We  develop  a probability  sum  rule,  like  that of  the  electronic
structure      case      \cite{Boykin_JPCM_2007,      Boykin_PRB_2005,
Boykin_PRB_2007} which leads to  a method for band determination. The sum
of the  square  magnitudes of  the  $m$-th  components (corresponding  to
$\vec{q}_m =  \vec{Q} +  \vec{G}_m$) of the  $C_p^{(\alpha, w)}(\vec{Q})$
over atoms and components of motion is the projection probability for the
SC state $p$ onto the PC states of wavevector $\vec{q}_m$:
\begin{align}
\label{eq_sumrule1}
\begin{split}
\mathcal{P}(E_p, \vec{q}_m) 
= \sum_{\alpha, w}\left|\Big[\vec{C}_p^{(\alpha, w)}(\vec{Q})\Big]_m\right|^2 = \\
\sum_{s'=1}^{3r}\sum_{s=1}^{3r}a_{p; (m,s')}^* a_{p; (m,s)} \times \\
\Bigg[ \sum_{\alpha, w} b_{s'}^{(\alpha,w)*}(\vec{Q} + \vec{G}_m)
                     b_{s}^{(\alpha,w)}(\vec{Q} + \vec{G}_m) \Bigg] \\
= \sum_{s=1}^{3r}|a_{p; (m,s)}|^2
\end{split}
\end{align}
where the last step follows from  the fact that the quantity in square
brackets is the  inner product $\vec{b}_{s'}^{\dagger} \cdot \vec{b}_s
=  \delta_{s',s}$. Next  sum eq.  (\ref{eq_sumrule1}) over  SC states,
$p$, and  replace the $a_{p; (m,s)}$ using  the orthogonality relation
$\vec{u}_{s',j}^{\dagger}(\vec{Q}'      +      \vec{G}_{m'})     \cdot
\vec{u}_{s,j}^{\dagger}(\vec{Q}  +   \vec{G}_{m})  =  \delta_{\vec{Q},
\vec{Q}'} \delta_{m,m'} \delta_{s,s'}$ and
\begin{align}
\label{eq_sumrule2}
\begin{split}
\vec{u}_{s,j}^{\dagger}(\vec{Q} + \vec{G}_m) \cdot \vec{V}_{p,j}  &= \\
       \sum_{s'=1}^{3r} \sum_{m'=1}^{N_c} 
       a_{p; (m',s')} \vec{u}_{s,j}^{\dagger}(\vec{Q} + \vec{G}_m) \cdot 
                  \vec{u}_{s',j}(\vec{Q} + \vec{G}_{m'}) &= \\
       a_{p; (m,s)} 
\end{split}
\end{align}
twice. Making the replacement in eq. (\ref{eq_sumrule1}) results in
\begin{align}
\label{eq_sumrule3}
\begin{split}
\sum_{p=1}^{3rN_c}\sum_{\alpha=1}^{r}\sum_{w=x}^{z} 
\left|\Big[\vec{C}_p^{(\alpha, w)}(\vec{Q})\Big]_m\right|^2 = 
\sum_{s=1}^{3r} \vec{u}_{s,j}^{\dagger}(\vec{Q} + \vec{G}_m) \cdot \\
                \left[ \sum_{p=1}^{3rN_c} 
                     \vec{V}_{p,j} \cdot \vec{V}_{p,j}^{\dagger} 
                \right] \cdot
                \vec{u}_{s,j}(\vec{Q} + \vec{G}_m).
\end{split}
\end{align}
The sum in  square brackets is nothing more  than the closure relation
for the eigenvectors of an Hermitian matrix,
\begin{align}
\label{eq_closure}
\sum_{p=1}^{3rN_c}\vec{V}_{p,j} \cdot \vec{V}_{p,j}^{\dagger} = \mat{1}_{3rN_c}
\end{align}
so that eq. \ref{eq_sumrule3} becomes the probability sum rule:
\begin{align}
\label{eq_sumrulefinal}
\begin{split}
\sum_{p=1}^{3rN_c}\sum_{\alpha=1}^{r}\sum_{w=x}^{z} 
\left|\Big[\vec{C}_p^{(\alpha, w)}(\vec{Q})\Big]_m\right|^2 & = \\
\sum_{s=1}^{3r} \vec{u}_{s,j}^{\dagger}(\vec{Q} + \vec{G}_m) \cdot 
              \vec{u}_{s,j}(\vec{Q} + \vec{G}_m) & = \\
\sum_{s=1}^{3r} \delta_{s,s} & = 3r.
\end{split}
\end{align}
In eq.  (\ref{eq_sumrulefinal}) we immediately recognize  that $3r$ is
the total number of bands at each $\vec{q}_m = \vec{Q} + \vec{G}_m$.

The  sum  rule  eq.  (\ref{eq_sumrulefinal})  suggests  the  following
general approach  for determining approximate  band positions: Compute
the  cumulative  probability for  SC  energies  $E_p  < E_B$  (the  SC
energies  are  in ascending  order)  at  fixed  $\vec{q}_m =  \vec{Q}  +
\vec{G}_m$, $m = 1, \dots, N_c$,
\begin{align}
\label{eq_Ptot}
\begin{split}
\mathcal{P}_{cum}(E_B, \vec{Q} + \vec{G}_m) & = 
\sum_{p=1}^B \mathcal{P}(E_p, \vec{q}_m) \\ 
& = \sum_{p=1}^{B} \sum_{\alpha=1}^{r}\sum_{w=x}^{z}
\left|\Big[\vec{C}_p^{(\alpha, w)}(\vec{Q})\Big]_m\right|^2, \\
B & \in [1,3rN_c]
\end{split}
\end{align}
and look  for gaps. Whenever the cumulative  probability has increased
by  unity  with  increasing  energy  a band  has  been  crossed.  This
observation encapsulates  the essential physics of  the procedure, but
refinements are necessary to make it automated and practical.

The physics of  effective phonon bands differs from  that of effective
electron bands in a few  important respects. First, the actual or near
degeneracy of the optical modes  throughout most of the Brillouin zone
is generally  much stronger than  degeneracies in the  electron bands,
except  near high symmetry  points (e.g.,  the heavy-  and light-holes
near $\Gamma$).   Second, in a  tight-binding electronic bandstructure
model -- recall that its  unfolding is mathematically identical to the
phonon  case  --  variations   in  the  onsite  and  neighboring  atom
parameters are often only  moderate.  For the phonon problem, however,
replacing one atom with another from  the same column as happens in an
alloy  results  in  a  significant  mass  change.   Thus  the  spreads
(uncertainties) in  the phonon bands  can be relatively  large.  Taken
together,  these observations  led  us to  modify  our effective  band
determination   algorithm   from   the   electronic   structure   case
\cite{Boykin_JPCM_2007}.

\begin{figure}[t]
 \centering	
 \includegraphics[scale=0.8]{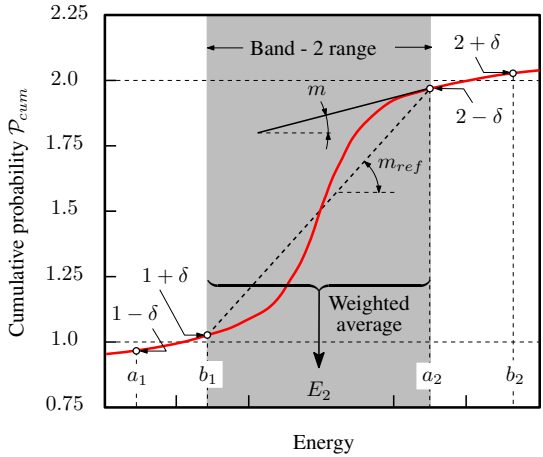}
 \caption{Example of the band  determination algorithm. Red solid line
(color  online):  The  cumulative probability,  $\mathcal{P}_{cum}$,
versus  energy,  shows  that  Band  2  falls  between  energies  $b_1$
($\mathcal{P}_{cum} = 1 + \delta$) and $a_2$ ($\mathcal{P}_{cum} = 2 -
\delta$), the grey  shaded region. The average slope  over the step in
cumulative probability  is $m_{ref}$ (black  doted line) and  the last
$5\%$ of $(E, \mathcal{P}_{cum})$ points  in the range are fitted to a
line with slope $m$ (solid black line). Because $m < m_{ref}$ the step
is  distinct and Band  2 can  be differentiated  from the  next higher
band(s).  The  probability-weighted  average  energy is  indicated  by
$E_2$.
}
 \label{fig_Fig1}
\end{figure}

In  the  modified method,  we  eliminate  the  parameters for  minimum
resolvable gap and minimum probability. Instead, we concentrate on the
cumulative  probability  and  its  slope. The  cumulative  probability
converges with supercell size (see Sec.  \ref{sec_Results} below) and,
as we  note in connection  with electron bands, step  determination is
simpler  than peak  determination  \cite{Boykin_JPCM_2007}.  Our  band
determination   method    is   given   below    and   illustrated   in
Fig.  \ref{fig_Fig1}, where  we  plot the  cumulative probability  for
fixed PC wavevector $\vec{q}_m = \vec{Q} + \vec{G}_m$ in the vicinity of
the second band edge for an hypothetical system. There are two control
parameters:   $\delta$  and   $\texttt{slopelim}$.  $\delta$   is  the
difference  in the  cumulative  probability from  an  integer used  to
bracket  integral values,  and $\texttt{slopelim}$  is  the cumulative
probability  slope above which  the current  candidate band  cannot be
separated  from the next  higher one.   In practice  we have  found to
$\delta = 0.05$ and $\texttt{slopelim}  = 1.0$ work well. The steps in
band determination are:
\begin{enumerate}
\item Bracket all integral values, $j$, of the cumulative probability,
denoted   by    the   energy    ranges   $[a_j,   b_j]$.    That   is,
$\mathcal{P}_{cum}(a_j,      \vec{q}_m)      =      j     -      \delta$,
$\mathcal{P}_{cum}(b_j,\vec{q}_m)  = j  + \delta$.   Fig. 1  shows these
brackets for $\mathcal{P}_{cum} = 1,  2$ . The band $(j+1)$ then falls
somewhere  between energies  $b_j$ and  $a_{j +  1}$, as  shown  in the
shaded  area of  Fig. 1  for  band 2.  Here band  2 is  nondegenerate;
degeneracies are treated in Step 3.
\item   Next determine whether  or not  the band  in the  range $[b_j,
a_{j+1}]$  can be  resolved  from the  next-higher band.   Physically,
resolution  is   not  possible  when  the  slope   of  the  cumulative
probability is  too large  near the  upper end of  the range:  A rapid
increase in the cumulative probability near the upper end of the range
means  that the  current band  and  the next  higher one  are for  all
practical purposes degenerate. We check the slope by fitting a line to
the last $5\%$  of points in the range  $[b_j, a_{j+1}]$ and comparing
it to the average slope over the entire interval, that of the straight
line connecting points  $(b_j, j+\delta)$ and $(a_{j+1}, j+1-\delta)$,
denoted $m_{ref}$. If
\begin{align}
\label{eq_mref}
\frac{m}{m_{ref}} > \texttt{slopelim}, 
m_{ref} = \frac{1 - 2\delta}{a_{j+1} - b_j}
\end{align}
the  current  band  cannot be  resolved  and  it  is merged  into  the
next-higher   band.     In   Fig.    \ref{fig_Fig1},    $m/m_{ref}   <
\texttt{slopelim}$  and therefore  band  2 can  be  resolved, and  its
energy is the indicated by the weighted average value, $E_2$.
\item Degneracies are characterized  by a zero-bracket: This situation
occurs when there is no  cumulative probability sample satisfying $j -
\delta \leq  \mathcal{P}_{cum} \leq j +  \delta$. (Due  to the finite
size of the SC the  cumulative probability is discrete.)  In this case
the candidate $j$-th band is merged into a doubly-degenerate band with
the  $(j+1)$-st and  the range  under consideration  is $[b_j,  a_{j +
2}]$.   This effective  band determination  method is  applied  to the
phonon bands of In$_x$Ga$_{1-x}$As in Sec. \ref{sec_Results} below.
\end{enumerate}

Once the bands  have been determined, the band  energies are computed.
Although   the  dynamical  matrix   eigenproblem  has   an  eigenvalue
$\omega^2$, or equivalently $E^2$,  we continue to compute the average
energy  as in the  electronic structure  case \cite{Boykin_JPCM_2007}.
Once the range  of SC energies contributing to the  $j$-th PC band has
been found  by the  procedure above,  the PC energy  for this  band is
computed as:
\begin{align}
\label{eq_Eavg}
\epsilon_j(\vec{Q} + \vec{G}_m) = 
\dfrac {\sum\limits_{i = M}^N \mathcal{P}(E_i, \vec{Q} + \vec{G}_m) \cdot E_i}
       {\sum\limits_{i = M}^N \mathcal{P}(E_i, \vec{Q} + \vec{G}_m)}
\end{align}
where the SC  states $i = M, M+1, \ldots, N$  contribute to the $j$-th
PC band. The energy range for  a set of degenerate bands is determined
by Step  3 above and  the set's average  energy is computed as  in the
electronic structure  case \cite{Boykin_JPCM_2007}.  We  note that for
strongly  peaked  functions eq.  (\ref{eq_Eavg})  and  a weighted  RMS
computation over  $E^2$ will give essentially the  same results.  More
importantly,  because the band  positions are  determined in  terms of
$E$, not $E^2$, eq.  (\ref{eq_Eavg}) is more fully consistent with the
band determination method.

\section{Results}
\label{sec_Results}

\begin{table}[t]
  \caption{Keating \cite{Keating_PhysRev_1966} parameters for InAs and
GaAs   in    both   Random    Alloy   and   VCA    calculations   from
Ref. \onlinecite{Pryor_JAP_1998}. Units are $N/m$.}
  \begin{ruledtabular}
    \begin{tabular}{ccc}
      &  
      $\alpha$  & 
      $\beta$  \\
      \colrule 

      InAs        & 
      $35.18$ & 
      $5.49$ \\
      
      GaAs  &
      $41.19$  &
      $8.94$ \\
    \end{tabular}
  \end{ruledtabular}
  \label{tab_Params}
\end{table}

We   demonstrate  the   effective  phonon   bandstructure   method  of
Sec.  \ref{sec_Method}  above  by  calculating the  phonon  bands  for
In$_x$Ga$_{1-x}$As     alloys      using     the     Keating     model
\cite{Keating_PhysRev_1966}.   The   parameters  for  InAs   and  GaAs
\cite{Pryor_JAP_1998}  are  listed  in  Table  \ref{tab_Params}.   The
Keating  model  has  deficiencies  \cite{Sui_PRB_1993,  Paul_JCE_2010,
Steiger_PRB_2011};   however   it   does  accurately   reproduce   the
longitudinal acoustic (LA)  mode from $\Gamma$ to L.   The bulk phonon
bands   for  GaAs   and   InAs  reproduced   by   the  Keating   model
\cite{Keating_PhysRev_1966} are included  in the supplemental material
\cite{Supplement}  for   this  paper.    In  the  Random   Alloy  (RA)
calculations we use the geometric average of the GaAs and InAs Keating
$\beta$  (bond-bending) parameters whenever  an As  atom is  the common
nearest-neighbor  to  both  a Ga  and  an  In  atom in  the  bond-pair
sum. Otherwise, we use the  appropriate bulk parameters for the single
bond     ($\alpha$)    or     bond-pair     ($\beta$).     For     the
In$_{0.5}$Ga$_{0.5}$As    Virtual    Crystal    Approximation    (VCA)
calculations used as  a basis for comparison, we  employ the geometric
average of the respective Keating parameters.

\begin{figure}[t]
 \centering	
 \includegraphics[scale=0.8]{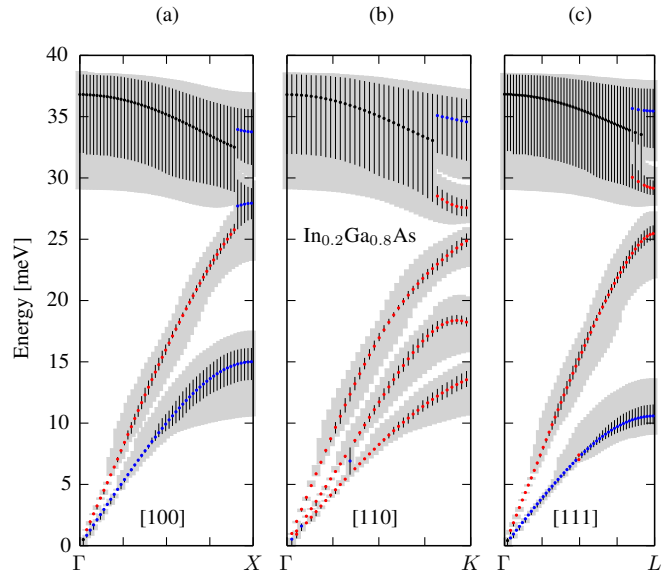}
 \caption{Random Alloy (RA)  unfolded bands for In$_{0.2}$Ga$_{0.8}$As
along $[100]$ (a), $[110]$ (b), and $[111]$ (c). Dots indicate average
energies and  dot color (online)  indicates degeneracy, $D$:  red (1),
blue (2), or black (3). Black lines and grey bars denote the spread in
probability for  the band they surround:  $0.25 \le \mathcal{P}_{band}
\le D  - 0.25$ (black) or  $0.05 \le \mathcal{P}_{band} \le  D - 0.05$
(grey).  }
 \label{fig_Fig2}
\end{figure}

\begin{figure}[t]
 \centering	
 \includegraphics[scale=0.8]{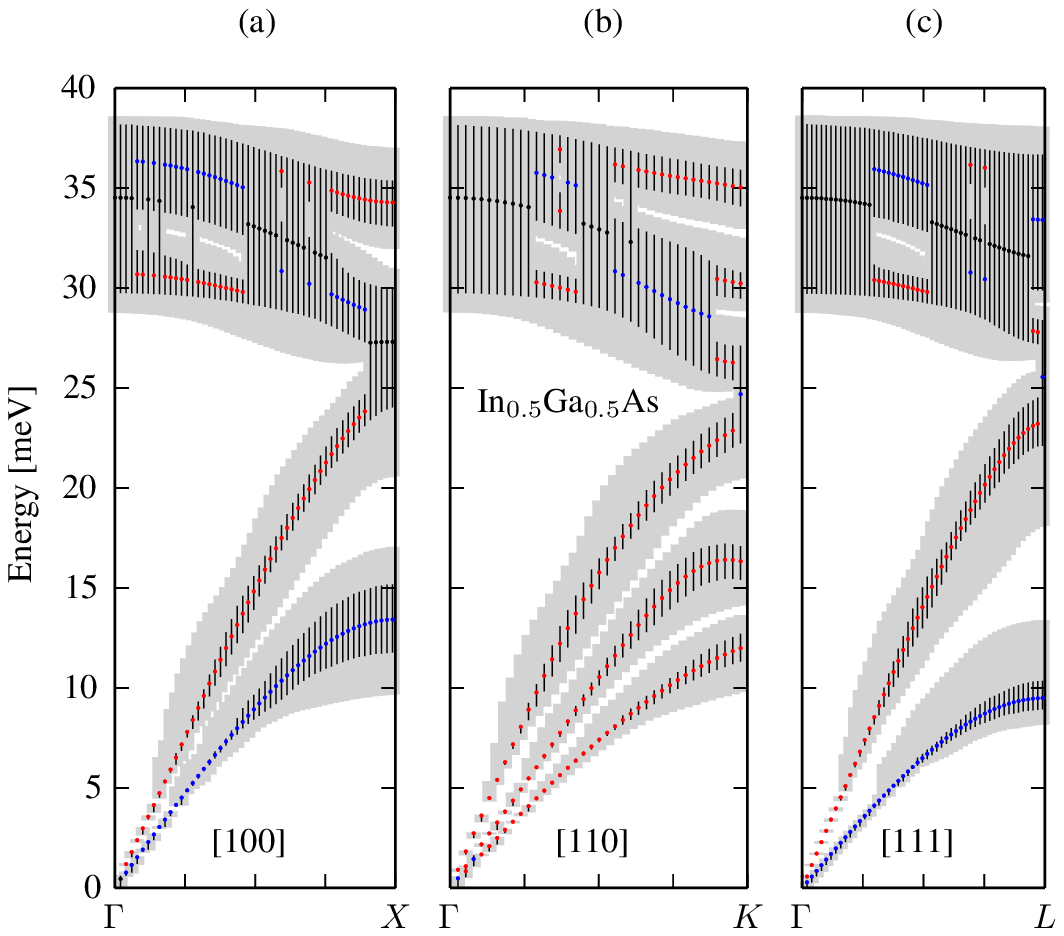}
 \caption{Random Alloy (RA)  unfolded bands for In$_{0.5}$Ga$_{0.5}$As
along $[100]$ (a), $[110]$ (b), and $[111]$ (c). The symbols are the
same as in Fig. \ref{fig_Fig2}.}
 \label{fig_Fig3}
\end{figure}

\begin{figure}[b]
 \centering	
 \includegraphics[scale=0.8]{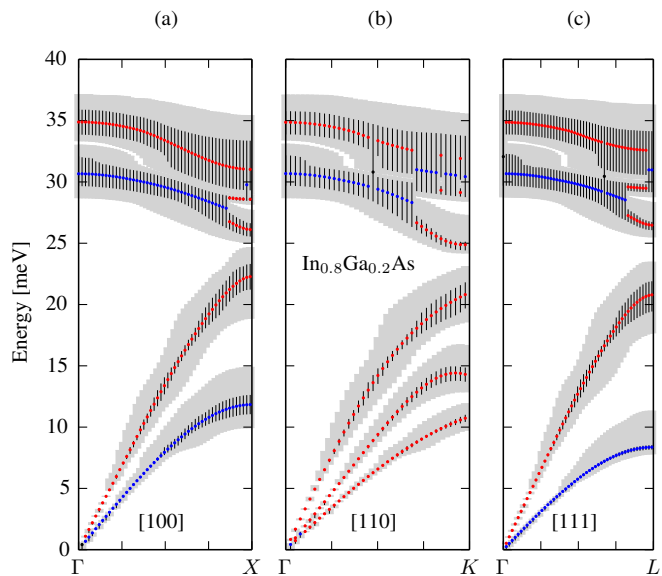}
 \caption{Random Alloy (RA)  unfolded bands for In$_{0.8}$Ga$_{0.2}$As
along $[100]$ (a), $[110]$ (b), and $[111]$ (c). The symbols are the
same as in Fig. \ref{fig_Fig2}.}
 \label{fig_Fig4}
\end{figure}

Figures   \ref{fig_Fig2}-   \ref{fig_Fig4}   show  the   RA   unfolded
In$_x$Ga$_{1-x}$As bands for  $x=0.2$, $0.5$, and $0.8$, respectively,
along each  of the symmetry directions $[100]$,  $[110]$, and $[111]$.
The   special   unit  cells   for   these   directions  (see   Section
\ref{sec_PrimitiveCells}) are  used and in each case,  $N_1 \times N_2
\times  N_3  =  101  \times  4  \times 4$.   As  discussed  below  the
bond-length  distribution for  this size  cell was  well-converged. In
these figures, dots indicate the weighted average energy and dot color
(online) denotes the degeneracy, $D$: red (1), blue (2), or black (3).
The  probability limits  on a  degenerate band  are best  expressed in
terms  of the band  probability: For  the band  falling in  the energy
range $[b_j  , a_{j + D}]$, $\mathcal{P}_{band}  = \mathcal{P}_{cum} -
j$. Black lines and grey bars denote the spread in probability for the
band   they  surround:   $0.25  \leq   \mathcal{P}_{band}  \leq   D  -
0.25$(black) or $0.05 \leq \mathcal{P}_{band} \leq D - 0.05$ (grey).

Generally, the  acoustic bands are  much better resolved than  are the
optical.  This development is not  surprising due to the fact that all
three optical  modes are very close together  throughout the Brillouin
Zone. Another  factor is  the large mass  discrepancy between  the two
cations involved, Ga and In. Large mass differences are an unavoidable
fact in  semiconductor alloys because the alloying  process results in
replacing  an atom  by a  different one  from the  same column  of the
periodic table. The effect on the  phonon bands near $q=0$ can be seen
in the simple two-atom-per-cell  chain model \cite{[{See, for example:
}]Harrison_Book_1979,  *Kittel_Book_1996}: $\omega_A =  \sqrt{1/(M_1 +
M_2)}$, $\omega_O  = \sqrt{(M_1  + M_2)/(M_1M_2)}$. Assuming  the same
force constant  for both materials  one finds $\Delta  \omega_A \Delta
\omega_O  \approx   0.3  (qa)$,  where   $\Delta  \omega_{\gamma}  =
\omega_{\gamma}^{\text{GaAs}}     -    \omega_{\gamma}^{\text{InAs}}$,
$\gamma \in \{A,O\}$.   Thus, there are good physical  reasons for the
greater spreads in the optical versus acoustic modes.

\begin{figure}[t]
 \centering	
 \includegraphics[scale=0.8]{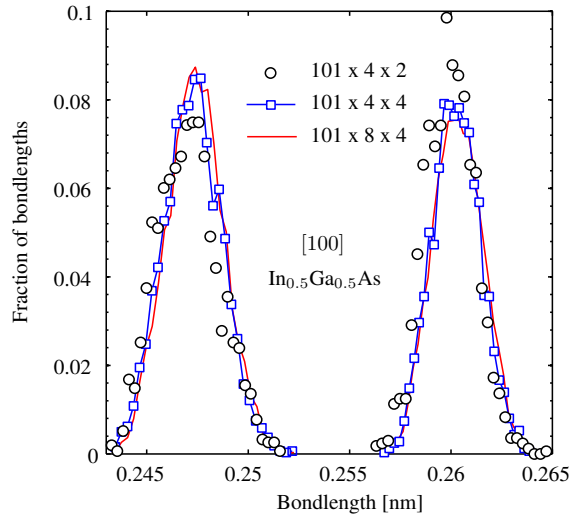}
 \caption{Bondlength     convergence    test    for     the    $[100]$
In$_{0.5}$Ga$_{0.5}$As  SCs (color  online). The  SCs use  the special
$[100]$ PC from Sec. II.B. Three different SC sizes are examined: $101
\times 4 \times 2$ (black  open circles), $101 \times 4 \times 4$(blue
solid line and  blue open squares), $101 \times  8 \times 4$(red solid
line).  The  two largest  cells agree well,  and therefore we  use the
$101  \times  4  \times  4$  SC  since it  affords  accuracy  at  less
computational cost than the $101 \times 8 \times 4$.}
 \label{fig_Fig5}
\end{figure}

We can gain additional insight into  the spreads of the alloy bands by
examining   the   eigenvectors    of   the   simple   two-atom   chain
model\cite{Kittel_Book_1996}.  At $q  =  0$, the  acoustic
branch  eigenvector is $[1/\sqrt{2},  1/\sqrt{2}]^T$. In  other words,
independent of  mass and force  constant the two  atomic displacements
have  equal magnitudes and  are in  phase. For  the optical  mode, the
displacements depend  on the masses and  are of opposite  sign (out of
phase).  This behavior  is clear  in the  RA calculations.  In  a like
manner, the greater  spreads near the Brillouin zone  boundary in both
the acoustic and  optical modes of the RA  calculations have parallels
in the simple  two-atom chain at $q = \pi/a$.   To make the discussion
concrete, assume that atom 1 is As while atom 2 is either Ga or In. In
the simple model  for $M_1 > M_2$ (GaAs) the  acoustic (A) and optical
(O)  mode  eigenvectors  are:  $\vec{u}_A  =  [1,0]^T$,  $\vec{u}_O  =
[0,1]^T$. In the acoustic mode  As is maximally displaced, while Ga is
at rest;  the optical mode is  the opposite.  For $M_1  < M_2$ (InAs),
these  results  are exchanged:  $\vec{u}_A  =  [0,1]^T$, $\vec{u}_O  =
[1,0]^T$, so  that for the  acoustic mode As  is stationary and  In is
maximally displaced, with the  optical mode the opposite.  Hence there
is a serious mismatch between these  two materials at $q = \pi/ a$ and
an increase in the band spread is hardly surprising.

The worst-case alloy,  In$_{0.5}$Ga$_{0.5}$As, is an optimal candidate
for  further  analysis.  Fig.  \ref{fig_Fig5}  shows  the  bond-length
distributions in  three different $[100]$ cells, $101  \times 4 \times
2$, $101 \times 4 \times 4$, $101\times 8 \times 4$.  It is clear that
the largest two  are nearly identical in terms  of bond lengths, while
the  smallest shows  significant deviations.   Thus,  the intermediate
cell,  $101 \times 4  \times 4$,  can safely  be used  for calculating
alloy   dispersions:  It  offers   good  accuracy   but  at   a  lower
computational cost.

\begin{figure}[t]
 \centering	
 \includegraphics[scale=0.8]{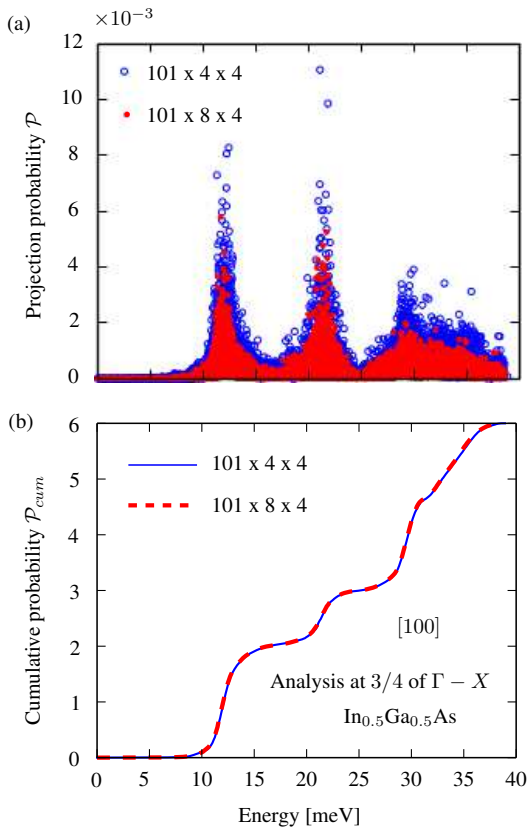}
 \caption{Projection  probability (a)  and cumulative  probability (b)
for the  $101 \times  4 \times  4$ (blue open  circles and  blue solid
line, color online) and $101  \times 8 \times 4$(red solid circles and
red dashed line) $[100]$ SCs $75\%$ of the way from $\Gamma$ to $X$ in
the PC first Brillouin  zone. The cumulative probability has obviously
converged, and the projection probability \emph{must} be lower for the larger
cell  because  the  sum  rule,  eq.  \ref{eq_sumrulefinal}  fixes  the
cumulative probability.  The larger  SC has twice  as many  samples so
each must contribute less due to the fixed cumulative probability.}
 \label{fig_Fig6}
\end{figure}

The reasons  for the  large uncertainties in  the phonon  bands become
clear  when  we  examine  the projection  probability  and  cumulative
probability  for  the  $[100]$  In$_{0.5}$Ga$_{0.5}$As $101  \times  4
\times    4$   supercell   at    a   specific    $\vec{q}$.    Figures
\ref{fig_Fig6}(a,b) show  these probabilities  $75\%$ of the  way from
$\Gamma$  to   $X$  in  the   PC  first  Brillouin  zone.   While  the
cumulative probabilities  for the two  cells are  essentially identical
the  projection  probabilities  differ.   Because the  sum  rule,  eq.
(\ref{eq_sumrulefinal}),   fixes  the   cumulative   probability,  the
projection probability  \emph{must} change when  the SC size changes:  In the
larger SC  there are  twice as many  probability samples so  each must
contribute less.  The twofold  degenerate Transverse Acoustic (TA) and
singly-degenerate  LA  modes  are  well  separated in  the  band  plot
Fig.  \ref{fig_Fig3}(a)  and this  fact  is  reflected  in both  Figs.
\ref{fig_Fig6}(a)  and \ref{fig_Fig6}(b).  In  Fig.  \ref{fig_Fig6}(a)
there are two  well-defined peaks corresponding to these  two modes at
around  $12$  meV (TA)  and  $22$  meV (LA).   In  a  like manner  the
cumulative  probability, Fig.   \ref{fig_Fig6}(b)  shows fairly  sharp
steps up  to $2$  between $10-15$  meV and up  to $3$  between $20-25$
meV.  Above $27$  meV  or so,  however,  Fig.  \ref{fig_Fig3}(a)  show
strong mixing of  all three optical modes, and  this mixing is obvious
in   both  Figs.    \ref{fig_Fig6}(a)   and  \ref{fig_Fig6}(b).    The
projection  probability, Fig.   \ref{fig_Fig6}(a), has  an ill-defined
clump from  around $27-38$ meV,  and the cumulative  probability, Fig.
\ref{fig_Fig6}(b)  has a  more  or less  continuous  rise from  around
$27-38$ meV with little evidence of a pronounced step.

\begin{figure}[t]
 \centering	
 \includegraphics[scale=0.8]{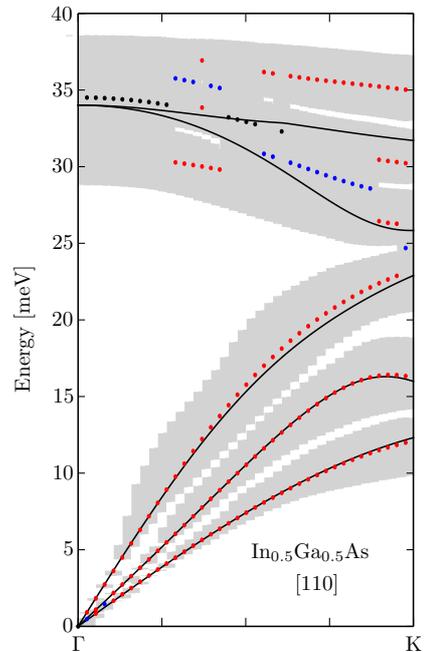}
 \caption{RA   and   VCA   bands  for   In$_{0.5}$Ga$_{0.5}$As   along
$[110]$. The RA bands are  represented by dots (color online) and grey
bars as in  Figs. \ref{fig_Fig2}-\ref{fig_Fig4}. Dots indicate average
energies and dot color indicates degeneracy, $D$:red (1), blue (2), or
black (3).  Grey bars  denote the spread  in probability for  the band
they surround:  $0.05 \le \mathcal{P}_{band} \le  D - 0.05$  . The VCA
bands are plotted with black solid lines.}
 \label{fig_Fig7}
\end{figure}

\begin{figure}[t]
 \centering	
 \includegraphics[scale=0.8]{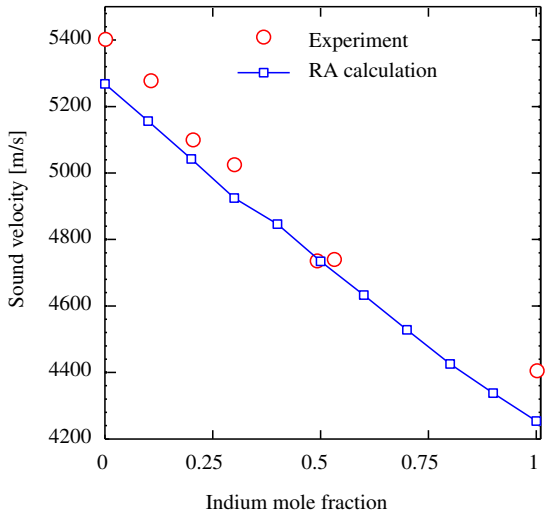}
 \caption{Sound   velocity  along   $[111]$   for  In$_x$Ga$_{1-x}$As,
computed from  $v_g = d\omega/dq$  at $q =  0$ for the LA  mode. Open
blue squares (color online) and blue  line (as a guide to the eye) are
the RA  unfolded results while  red open circles are  the experimental
data \cite{Wen_JAP_2006}.  There is very good agreement between the RA
results and experiment.
}
 \label{fig_Fig8}
\end{figure}

In Fig.  \ref{fig_Fig7} we compare  the VCA and RA bands along $[110]$
for  In$_{0.5}$Ga$_{0.5}$As.  The VCA  bands  are  plotted with  black
lines,   the   RA   results   with   dots  and   grey   bars   as   in
Figs. \ref{fig_Fig2}-\ref{fig_Fig4}. As  seen in \ref{fig_Fig3}(b) the
acoustic modes are  well-defined and the VCA in  fact agrees well with
the RA  results for these modes.   The RA optical  modes are generally
heavily mixed. Both trends have already been discussed with respect to
the $[100]$ bands.  Although  the one-dimensional model is perhaps not
quite so direct  an analogy in this case (the  planes have both anions
and cations, while  the $[100]$ are exclusively anion  or cation), the
optical modes are sufficiently close in energy that significant mixing
occurs. In contrast, the VCA  optical modes remain distinct because in
that case the crystal is perfectly ordered.

As mentioned above, the Keating model \cite{Keating_PhysRev_1966} does
accurately reproduce the acoustic modes  from $\Gamma$ to $L$, so that
a sound velocity calculation using RA  results based on it a good test
of the  effective phonon bandstructure model presented  here.  Fig.  8
shows the sound  velocity along $[111]$ (i.e., of  the LA mode) versus
In mole fraction: Open squares and  lines (to guide the eye) are the
RA   results    while   open   circles   are   experimental   results
\cite{Wen_JAP_2006}. The  $101 \times 4  \times 4$ $[111]$ SC  is used
for the RA  results, and the sound velocity $v_g =  dw/dq$ at $q=0$ is
calculated   with   a    forward-   difference   approximation.    The
uncertainties on  the RA  calculation are very  small so they  are not
shown: Note the tiny uncertainties for each $[111]$ LA mode near q = 0
in  Figures  \ref{fig_Fig2}-\ref{fig_Fig4}(c).   The  RA  calculations
match the experimental  results well with a maximum  relative error of
under $3\%$.  Better  agreement could be obtained by  using either the
modified valence-force-field (MVFF) \cite{Sui_PRB_1993, Paul_JCE_2010}
or  enhanced valence-force-field (EVFF)  \cite{Steiger_PRB_2011} models
instead of Keating's \cite{Keating_PhysRev_1966}.

\section{Conclusions}
\label{sec_Conclusions}
We have developed an effective phonon bandstructure calculation method
based  on Brillouin zone  unfolding.  As  in the  electronic structure
case  \cite{Boykin_JPCM_2007,  Boykin_PRB_2005,  Boykin_PRB_2007}  one
first randomly populates a SC with the atoms of an alloy in the proper
mole  fraction,  then  finds  the  SC eigenstates.   From  these,  one
projects out their contributions to  PC states of $\vec{q}_m = \vec{Q} +
\vec{G}_m$ . The probability sum  rule for the phonon problem leads to
an ansatz  for effective band  determination: Bands occur  at energies
where  the  cumulative  probability  makes integral  steps.   We  have
modified   the    band   determination   method\cite{Boykin_JPCM_2007,
Boykin_PRB_2005,  Boykin_PRB_2007}   to  better  align   it  with  the
different  physics of  the  vibrational spectrum  problem. Using  this
method  we  have  studied   the  effective  phonon  bandstructures  of
In$_x$Ga$_{1-x}$As alloys.  In general  we find that the optical modes
are  heavily  mixed whereas  the  acoustic  modes  tend to  be  better
defined.  These characteristics can be  at least partly explained by a
simple one-dimensional model \cite{Kittel_Book_1996}.  To validate the
effective phonon bandstructure method, we calculate the sound velocity
along $[111]$ versus  mole fraction and find very  good agreement with
experiment.  The  method developed here  should be useful  for thermal
problems in transistors, nanotransistors,  and other devices made from
semiconductor alloys.

\begin{acknowledgments}
This work was  supported in part by the Center  for Low Energy Systems
Technology  (LEAST), one of  six centers  of STARnet,  a Semiconductor
Research Corporation program sponsored by  MARCO and DARPA. The use of
nanoHUB.org  computational  resources  operated  by  the  Network  for
Computational  Nanotechnology  funded   by  the  US  National  Science
Foundation   under   grant   EEC-1227110,  EEC-0228390,   EEC-0634750,
OCI-0438246, and OCI-0721680 is gratefully acknowledged.
\end{acknowledgments} 
\vspace{1cm}

\appendix*
\section{}
The special PCs of Sec. \ref{sec_PrimitiveCells} are chosen so as to
probe  specific parts  of the  PC  Brillouin zone  when starting  from
$\vec{q} = \vec{0}$.  For all three cells, we follow the path $\vec{q}
:   \vec{0}  \rightarrow   (1/2)  \vec{\beta}_1^{[lmn]}$   (see  Table
\ref{tab_Cells}).   For   the  $[100]$   PC,  this  path   is  $\Gamma
\rightarrow X$,  while for the  $[111]$ PC, it is  $\Gamma \rightarrow
L$.  For  the $[110]$  cell,  this  path  crosses the  Brillouin  zone
boundary at $(3/8)  \vec{\beta}_1^{[110]}$, so the first three-fourths
of the path corresponds to  $\Gamma \rightarrow K$.  The last quarter,
$\vec{q}:     (3/8)     \vec{\beta}_1^{[110]}    \rightarrow     (1/2)
\vec{\beta}_1^{[110]}$ is easily shifted back into the first Brillouin
zone by adding  the reciprocal lattice vector $\vec{\beta}_2^{[110]}$.
After shifting, the  remainder now traverses the top  diamond from the
side midpoint to its center, $U \rightarrow X$ . Here we only plot the
bands for the  $\Gamma \rightarrow K$ portion of the  path since it is
of the most interest.

\clearpage
\pagebreak

\setcounter{equation}{0}
\setcounter{figure}{0}
\setcounter{table}{0}
\setcounter{page}{1}

\widetext
\begin{center}
\textbf{\large Supplemental  Material :  Brillouin zone  unfolding  method for
effective phonon spectra} 
\end{center}

\noindent\large{\textbf{1. Keating model for InAs and GaAs}}

\begin{figure}[!h]
 \centering	
 \includegraphics[scale=1]{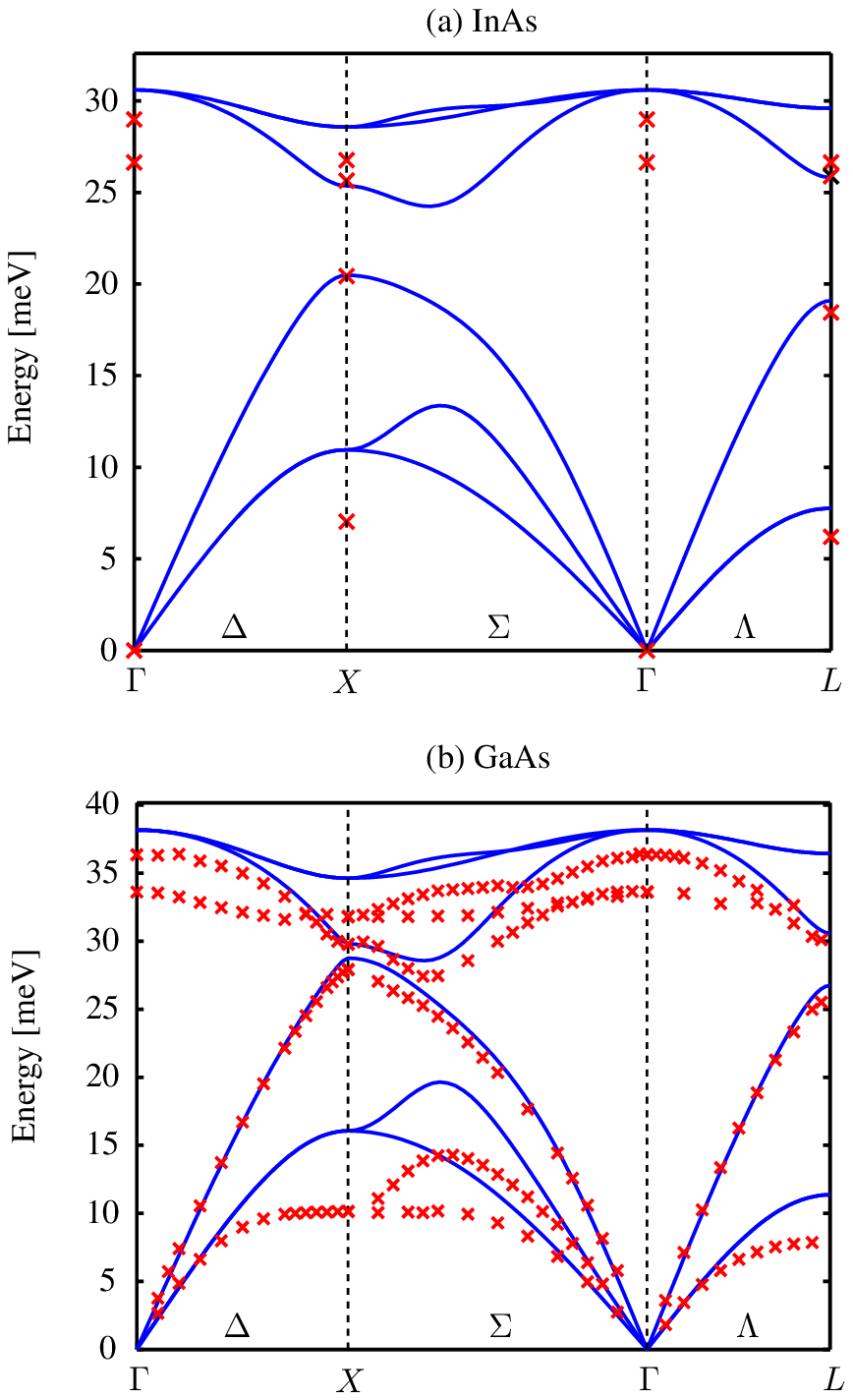}
 \caption{Comparison   of  the  Keating   model  (solid   lines,  with
parameters given in Table II of the main paper) with experimental data
(crosses, from Ref. \onlinecite{S_InAs} for InAs and Ref. 
\onlinecite{S_Strauch_JPCM_1990} for GaAs) for the phonon
dispersion  of (a)  InAs and  (b) GaAs.  Note that  the  Keating model
captures the LA mode in both materials well.}
\end{figure}

\noindent\large{\textbf{2. Convergence with supercell size}}\\
\newline
Fig.   6 of the  main paper  shows the  convergence of  the cumulative
probability for the  $101 \times 4 \times 4$ and  $101 \times 8 \times
4$  supercells  at  a point  $75  \%$  of  the distance  from  $\Gamma
\rightarrow X$.  Fig. 2  below shows the final effective bandstructure
obtained from these  two supercells. Note the good  convergence of the
effective  bandstructure with  supercell size.   The  differences seen
w.r.t the position of the mean and the spread in the optical bands are
artifacts of the  slope condition (eq.  28) of  the band determination
algorithm. As pointed out in the discussion connected to Fig. 6 of the
main paper,  the optical  bands are strongly  mixed. Hence,  there are
cases where  the slope  condition is just  about satisfied.   In these
cases, even  small differences in cumulative probability  with lead to
different binning  of energies, and their consequent  mean and spread.
Nevertheless, note that the  degeneracies are reported in a consistent
manner.\\

\begin{figure}[!b]
 \centering	
 \includegraphics[scale=1]{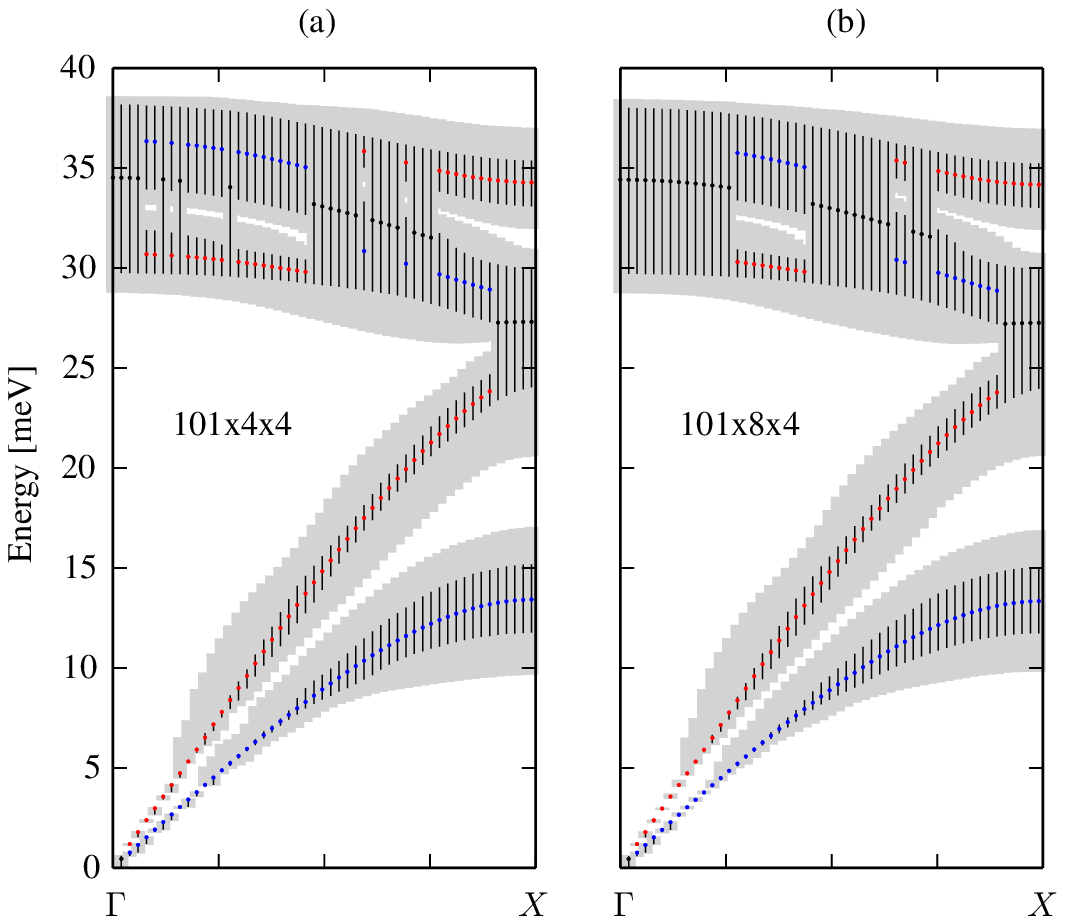}
 \caption{Convergence of the effective bandstructure with supercell size, shown
for In$_{0.5}$Ga$_{0.5}$As along the $[100]$ direction. The symbols are the same
as in Fig. 2 of the main paper.}
\end{figure}

\noindent\large{\textbf{3. Comparison of effective bandstructure along
equivalent directions in the Brillouin zone}}\\
\newline
A  perfect  crystal can  have  several  equivalent  directions in  the
Brillouin zone, based  on its symmetry. By definition,  a random alloy
has  no such  equivalent directions.  Nevertheless, if  the individual
constituents of an  alloy belong to the same  symmetry class (like for
example,  InAs  and  GaAs),  we  would  like  the  effective  unfolded
bandstructure to manifest  this symmetry. Ref. \onlinecite{S_Zunger_PRB_2012}
achieves  this  by averaging  the  cumulative  probability over  these
equivalent   directions,   prior    to   constructing   an   effective
bandstructure.   We  do  not  perform  this averaging  in  this  work.
However, we do not expect the  final result of such an averaging to be
significantly  different from  the results  obtained  from considering
only one of the many  equivalent directions. For example, Fig. 3 above
shows the effective  bandstructure of In$_{0.5}$Ga$_{0.5}$As along the
$[111]$ and  $[1\bar{1}1]$ directions, computed using   $101 \times 4
\times  4$ supercells.   The  supercell for  the  $[111]$ direction  is
constructed as  before, using the primitve cell  lattice vectors given
in  Table I of  the main  paper. The  supercell for  the $[1\bar{1}1]$
direction  uses $\alpha_1 =  (1,0,1), \alpha_2  = (0,1,1),  \alpha_3 =
(-1,-1,0)$ (specified  as earlier, in cartesian  coordinates and units
of $a/2$). The effective bandstructure indeed looks very similar for these
directions (the small differences w.r.t mean and spread in the optical
bands are due to the reason described in the previous section). \\

\begin{figure}[!t]
 \centering	
 \includegraphics[scale=1]{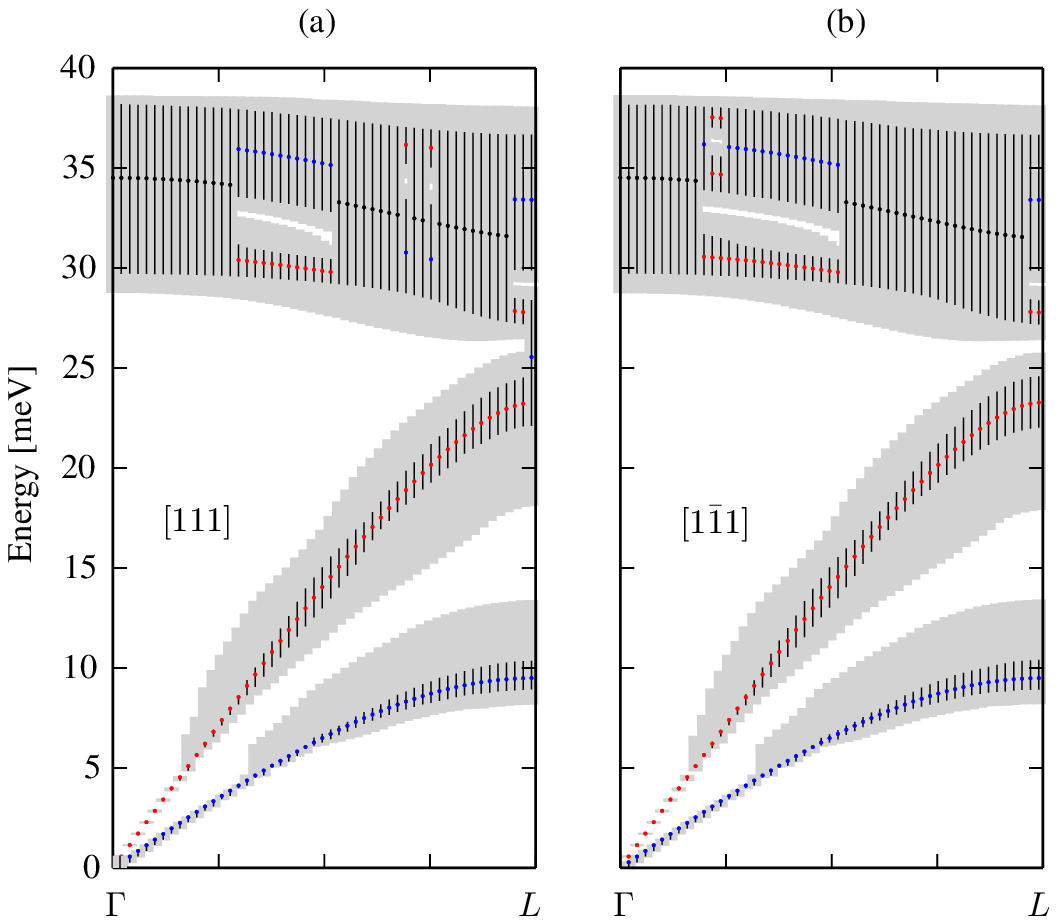}
 \caption{Comparison of  the effective bandstructure  along equivalent
directions  in the  Brillouin zone,  shown  for In$_{0.5}$Ga$_{0.5}$As
along  the  $[111]$ and  $[1\bar{1}1]$  directions. Calculations  were
performed using   $101 \times 4 \times 4$  supercells. The symbols are
the same as in Fig. 2 of the main paper.}
\end{figure}

\noindent\large{\textbf{4. Effective bandstructure of In$_{x}$Ga$_{1-x}$As 
for  $x = \mathbf{0.0, 0.1, \ldots 1.0}$}}\\
\newline
For   completeness,   we  present   the   computed  bandstructure   of
In$_{x}$Ga$_{1-x}$As  for $x  = 0.0,  0.1,$ $ \ldots  1.0$ in  Figs. 4-14
below. Note  that Figs.  2-4 of  the main paper  present the effective
phonon bandstructure of In$_{0.2}$Ga$_{0.8}$As, In$_{0.5}$Ga$_{0.5}$As and
In$_{0.8}$Ga$_{0.2}$As respectively, but are nevertheless repeated here.
All computations used the $101 \times 4 \times 4$ supercell.

\begin{figure}[!h]
 \centering	
 \includegraphics[scale=1]{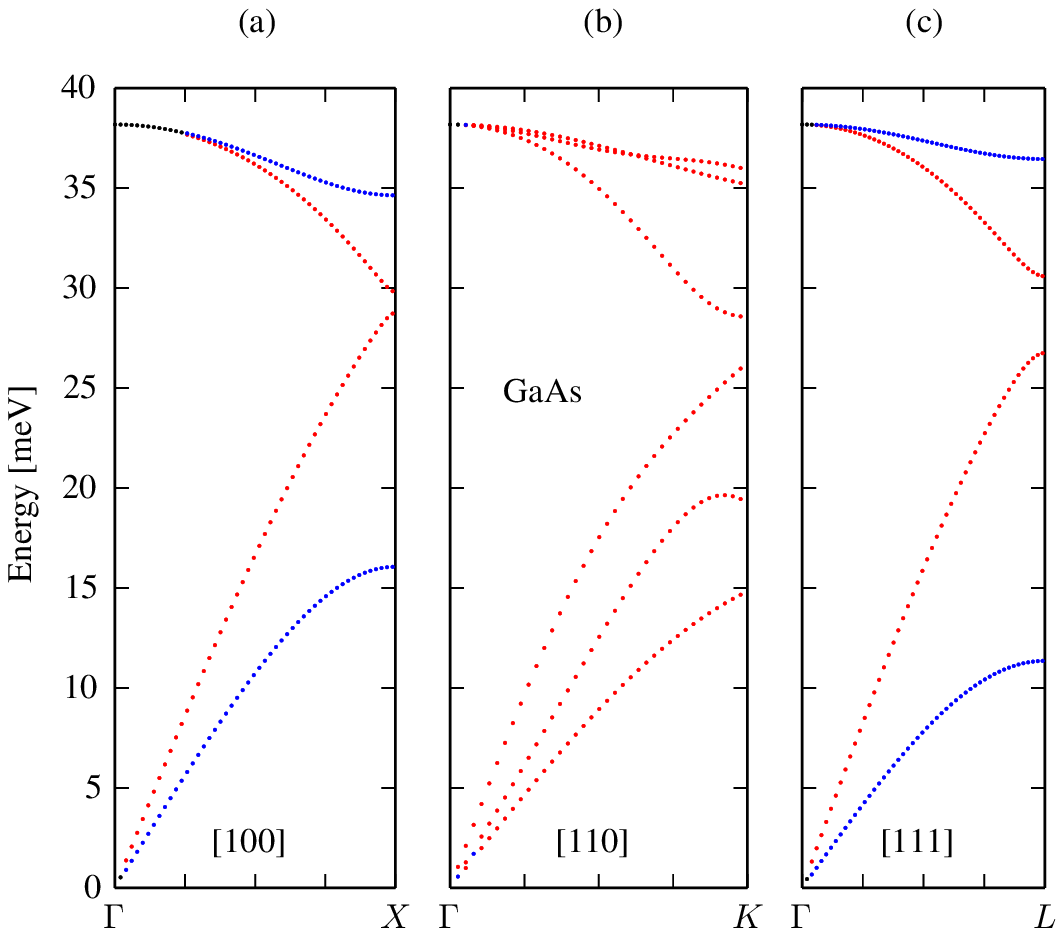}
  \caption{Unfolded bands for GaAs
along $[100]$ (a), $[110]$ (b), and $[111]$ (c). The symbols are the
same as in Fig. 2 of the main paper.}
\end{figure}

\begin{figure}[!h]
 \centering	
 \includegraphics[scale=1]{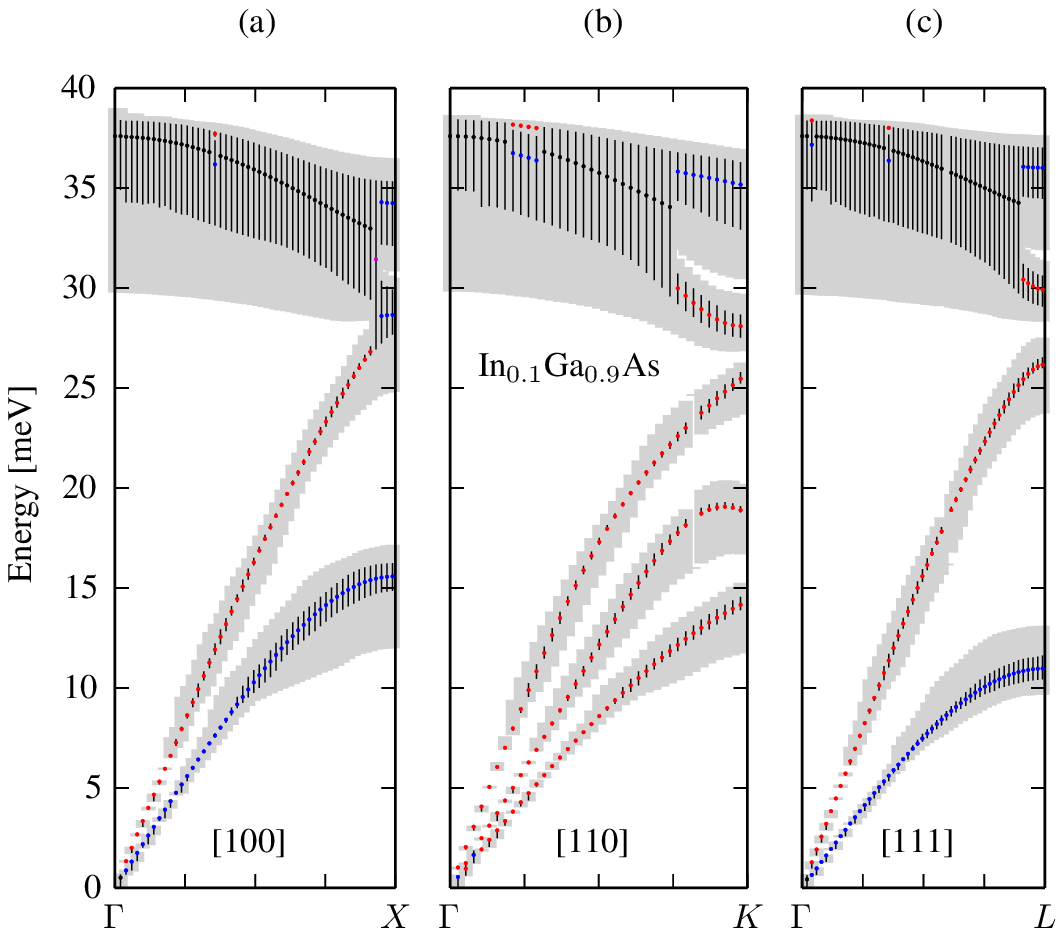}
  \caption{Random Alloy (RA)  unfolded bands for In$_{0.1}$Ga$_{0.9}$As
along $[100]$ (a), $[110]$ (b), and $[111]$ (c). The symbols are the
same as in Fig. 2 of the main paper.}
\end{figure}

\begin{figure}[!h]
 \centering	
 \includegraphics[scale=1]{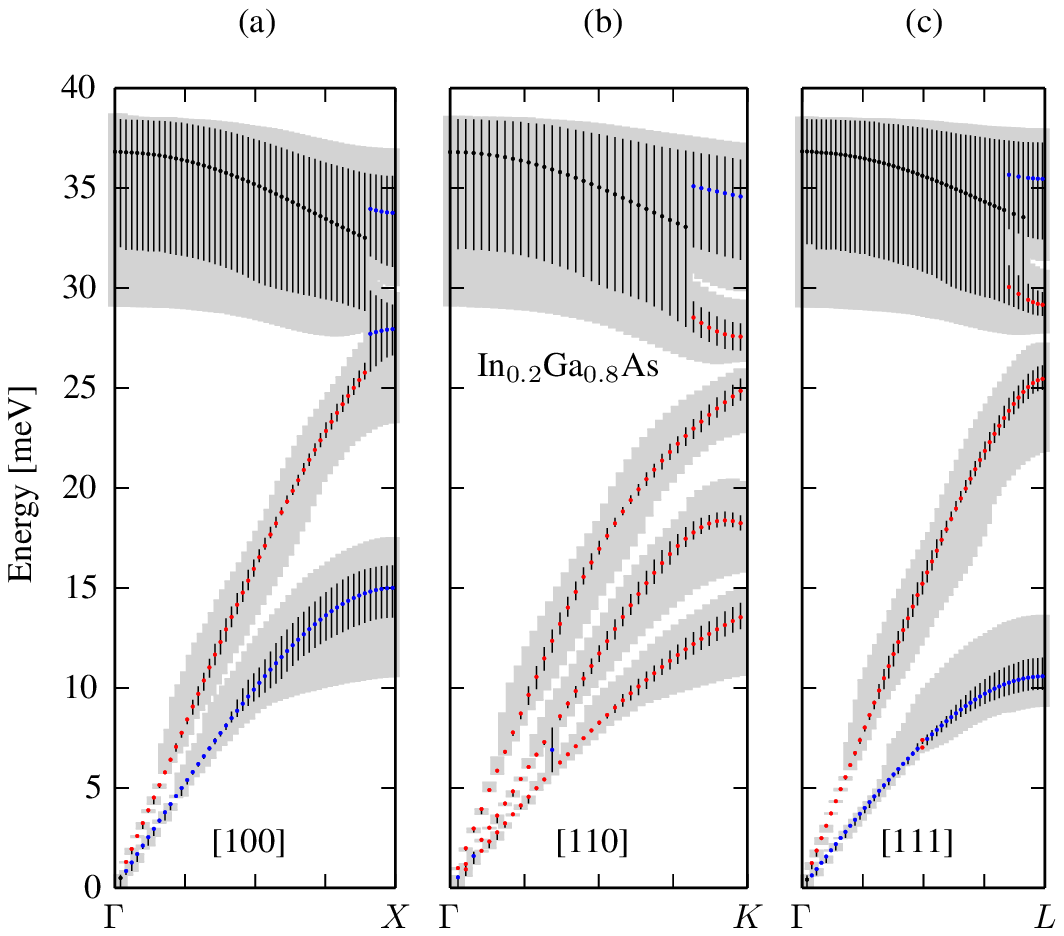}
  \caption{Random Alloy (RA)  unfolded bands for In$_{0.2}$Ga$_{0.8}$As
along $[100]$ (a), $[110]$ (b), and $[111]$ (c). The symbols are the
same as in Fig. 2 of the main paper.}
\end{figure}

\begin{figure}[!h]
 \centering	
 \includegraphics[scale=1]{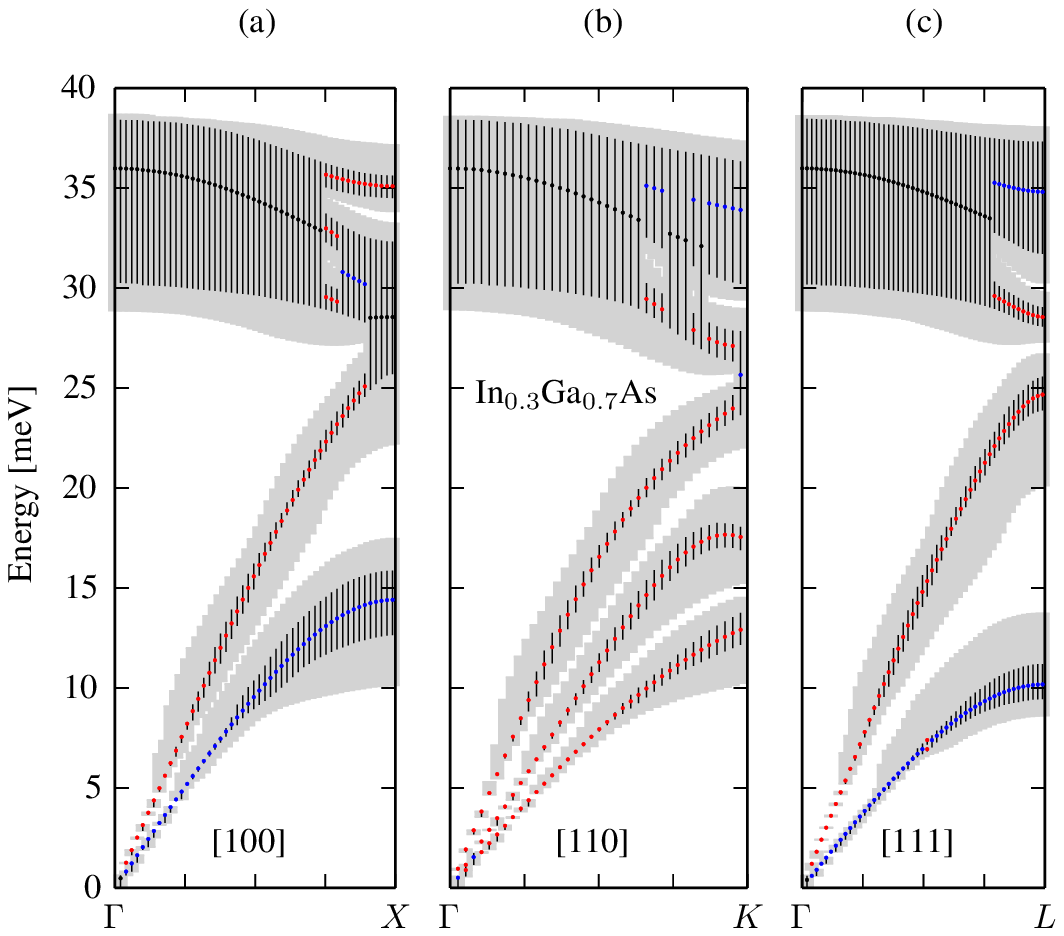}
  \caption{Random Alloy (RA)  unfolded bands for In$_{0.3}$Ga$_{0.7}$As
along $[100]$ (a), $[110]$ (b), and $[111]$ (c). The symbols are the
same as in Fig. 2 of the main paper.}
\end{figure}

\begin{figure}[!h]
 \centering	
 \includegraphics[scale=1]{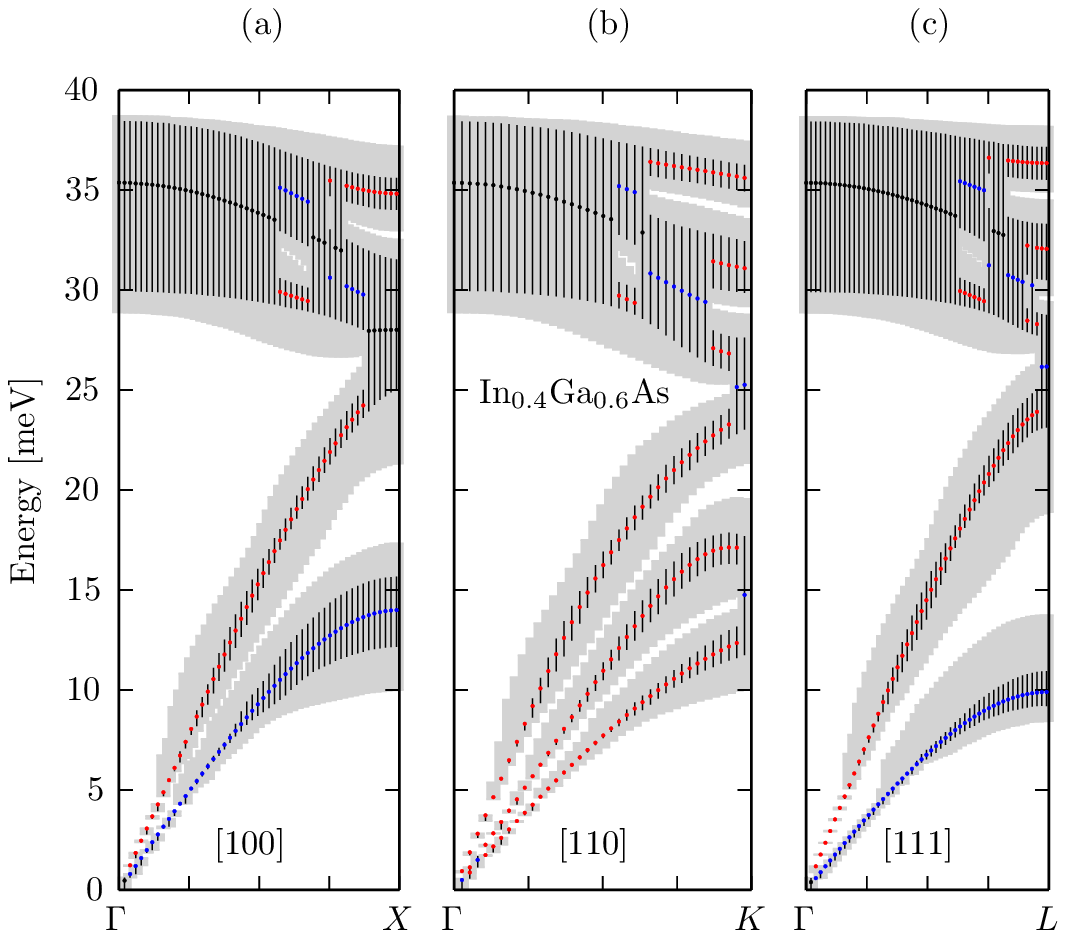}
  \caption{Random Alloy (RA)  unfolded bands for In$_{0.4}$Ga$_{0.6}$As
along $[100]$ (a), $[110]$ (b), and $[111]$ (c). The symbols are the
same as in Fig. 2 of the main paper.}
\end{figure}

\begin{figure}[!h]
 \centering	
 \includegraphics[scale=1]{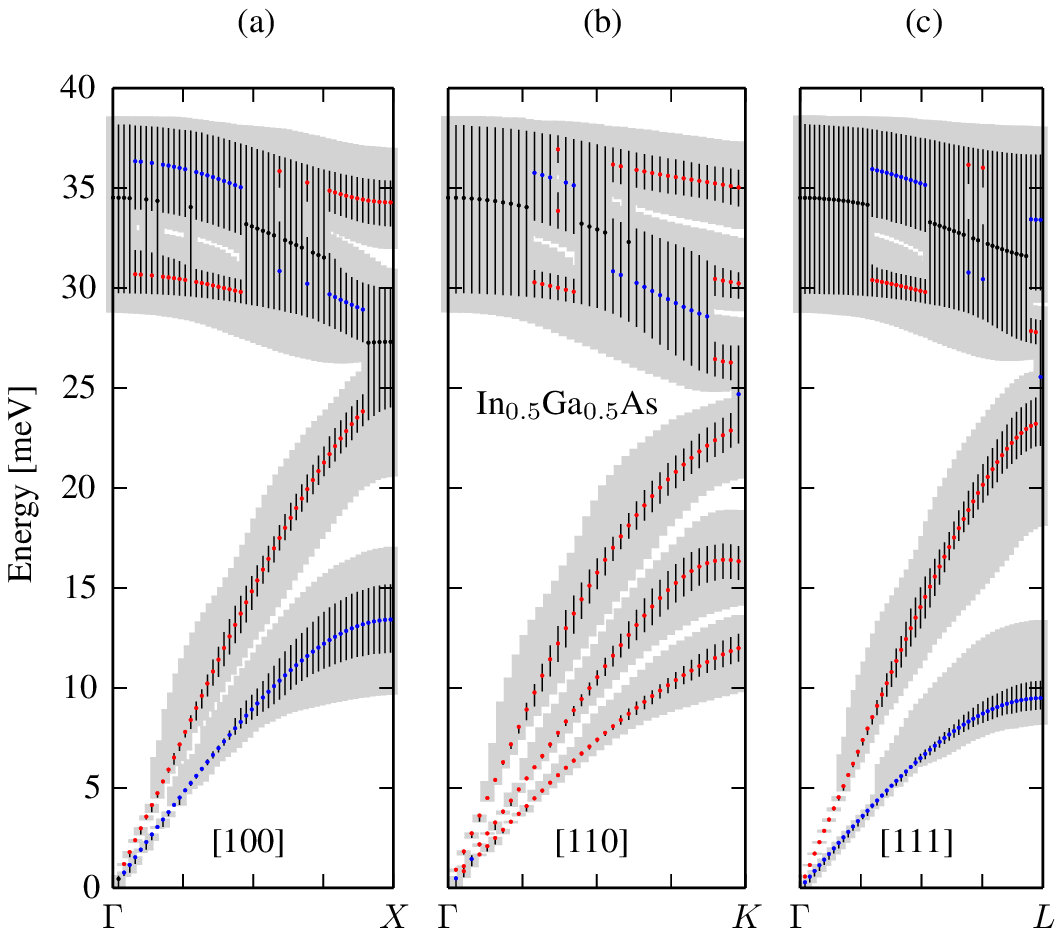}
  \caption{Random Alloy (RA)  unfolded bands for In$_{0.5}$Ga$_{0.5}$As
along $[100]$ (a), $[110]$ (b), and $[111]$ (c). The symbols are the
same as in Fig. 2 of the main paper.}
\end{figure}

\begin{figure}[!h]
 \centering	
 \includegraphics[scale=1]{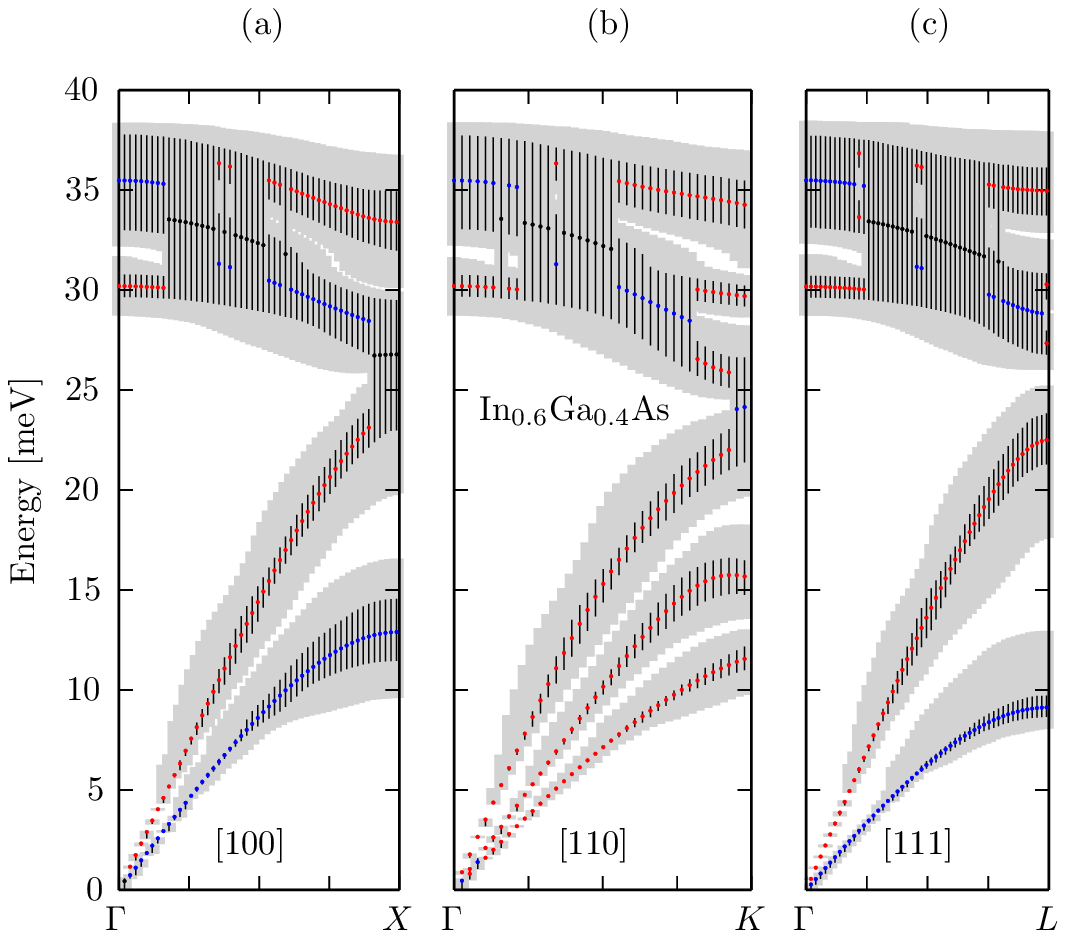}
  \caption{Random Alloy (RA)  unfolded bands for In$_{0.6}$Ga$_{0.4}$As
along $[100]$ (a), $[110]$ (b), and $[111]$ (c). The symbols are the
same as in Fig. 2 of the main paper.}
\end{figure}

\begin{figure}[!h]
 \centering	
 \includegraphics[scale=1]{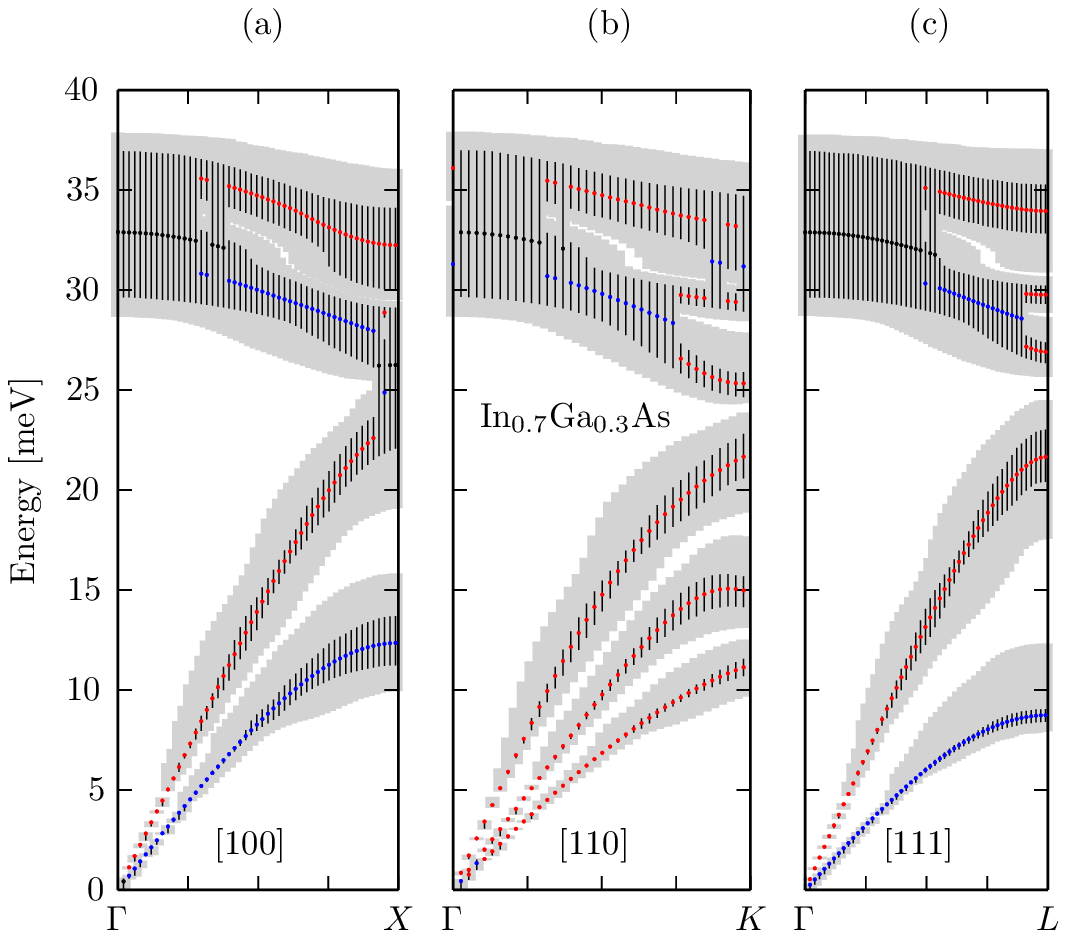}
  \caption{Random Alloy (RA)  unfolded bands for In$_{0.7}$Ga$_{0.3}$As
along $[100]$ (a), $[110]$ (b), and $[111]$ (c). The symbols are the
same as in Fig. 2 of the main paper.}
\end{figure}

\begin{figure}[!h]
 \centering	
 \includegraphics[scale=1]{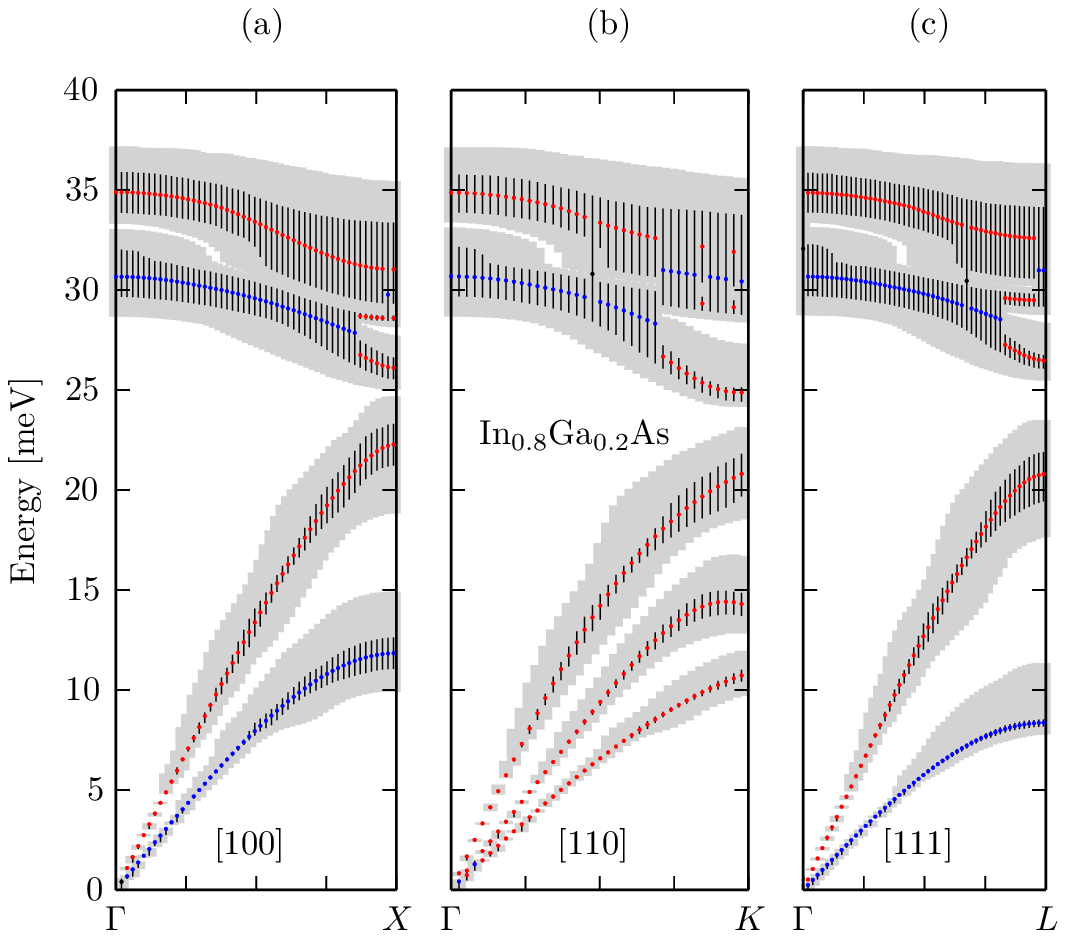}
  \caption{Random Alloy (RA)  unfolded bands for In$_{0.8}$Ga$_{0.2}$As
along $[100]$ (a), $[110]$ (b), and $[111]$ (c). The symbols are the
same as in Fig. 2 of the main paper.}
\end{figure}

\begin{figure}[!h]
 \centering	
 \includegraphics[scale=1]{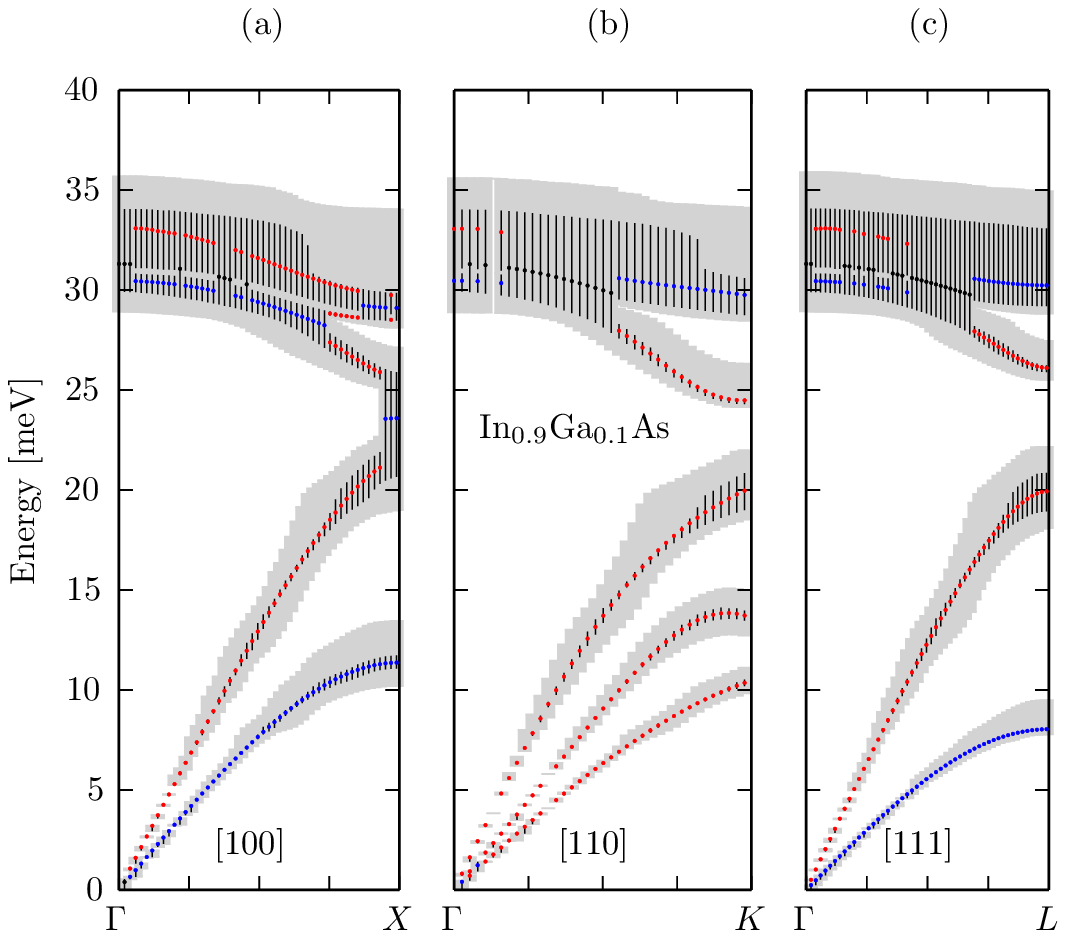}
  \caption{Random Alloy (RA)  unfolded bands for In$_{0.9}$Ga$_{0.1}$As
along $[100]$ (a), $[110]$ (b), and $[111]$ (c). The symbols are the
same as in Fig. 2 of the main paper.}
\end{figure}

\begin{figure}[!h]
 \centering	
 \includegraphics[scale=1]{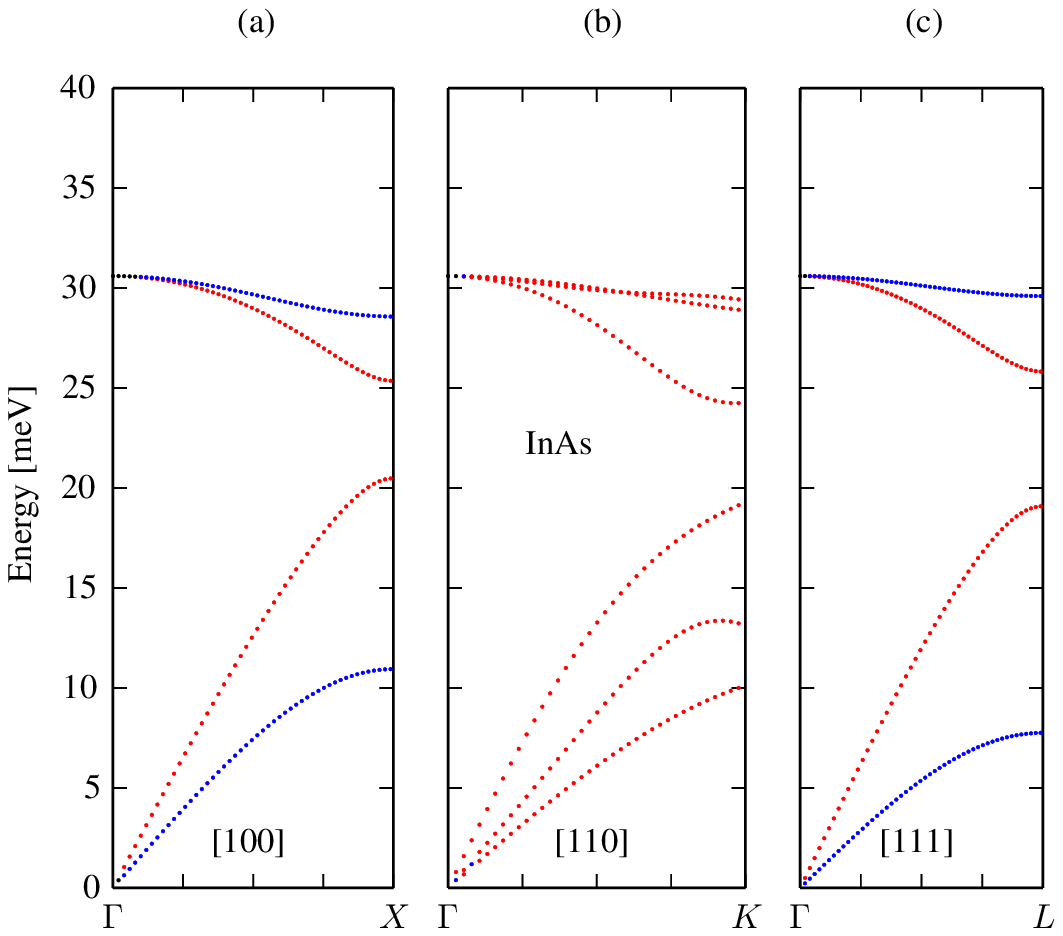}
  \caption{Unfolded bands for InAs
along $[100]$ (a), $[110]$ (b), and $[111]$ (c). The symbols are the
same as in Fig. 2 of the main paper.}
\end{figure}

\end{document}